\documentclass[review,authoryear,12pt]{elsarticle}
\usepackage{epsfig,verbatim}
\usepackage{color}
\usepackage{natbib,epsfig,graphicx,setspace,}
\usepackage{amsmath,amsthm,amssymb,enumerate}
\usepackage{mathrsfs}
\usepackage{bm,verbatim,subfigure}
\usepackage{booktabs}
\usepackage{color, xcolor}
\usepackage{booktabs}
\usepackage{rotating,epstopdf}
\makeatletter
\def\singlespace{\def\baselinestretch{1}\@normalsize}

\@addtoreset{equation}{section}

\renewcommand{\hat}{\widehat}
\def\singlespace{\def\baselinestretch{1}\@normalsize}

\makeatother

\newcommand{\bx}{\mbox{\bf x}}

\newcommand{\bp}{\mbox{\bf p}}

\newcommand{\bbeta}{\mbox{\boldmath$\beta$}}

\newcommand{\btheta}{\mbox{\boldmath$\theta$}}

\def\today{\ifcase\month\or
  January\or February\or March\or April\or May\or June\or
  July\or August\or September\or October\or November\or December\fi
  \space\number\day, \number\year}

\newdimen\biblioindent    \biblioindent=30pt

 at 7truept

\newcommand{\be}{\begin{equation}}
\newcommand{\ee}{\end{equation}}
\newcommand{\beq}{\begin{equation}}
\newcommand{\eeq}{\end{equation}}
\newcommand{\beqn}{\begin{eqnarray}}
\newcommand{\eeqn}{\end{eqnarray}}
\newcommand{\beqnn}{\begin{eqnarray*}}
\newcommand{\eeqnn}{\end{eqnarray*}}

\newtheorem{thm}{Theorem}[section]

\def\boxit#1{\vbox{\hrule\hbox{\vrule\kern6pt\vbox{\kern6pt#1\kern6pt}\kern6pt\vrule}\hrule}}

\begin{document}
\renewcommand{\baselinestretch}{1.0}
\begin{frontmatter}
\title{Robust Model Selection for Finite Mixture of Regression Models Through Trimming}
\author{
Sijia Xiang\footnote{School of Data Sciences, Zhejiang University of Finance and Economics, Hangzhou, P.R.China.
 Email: sjxiang@zufe.edu.cn. Xiang's research is supported by the National Nature Science Foundation of China grant 11601477.}\hspace{0.2cm}and\hspace{0.2cm} Weixin Yao\footnote{Department of Statistics, University of California, Riverside, CA 92521, USA. Email: weixin.yao@ucr.edu. Yao's research is supported by NSF grant DMS-1461677 and Department of Energy with the award No: 10006272.}
}


\begin{abstract}
In this article, we introduce a new variable selection technique through trimming for finite mixture of regression models. Compared to the traditional variable selection techniques, the new method is robust and not sensitive to outliers. The estimation algorithm is introduced and numerical studies are conducted to examine the finite sample performance of the proposed procedure and to compare it with other existing methods.
\end{abstract}


\end{frontmatter}
\pagestyle{plain}

\section{Introduction}
\setcounter{page}{1}
Finite mixture of regressions (FMR), also known as switching regression models in econometrics, has been widely used in scenarios when a single regression fails to adequately explain the relationship between the variables. Applications of it can be seen in econometrics (Wedel and DeSarbo, 1993; Fr$\ddot{u}$hwirth-Schnatter, 2001), in epidemiology (Green and Richardson, 2002), and also in outlier detection or robust regression estimation (Young and Hunter, 2010). See also, the book by McLachlan and Peel (2000) for a comprehensive review of finite mixture models, and the books by Skrondal and Rabe-Hesketh (2004) for applications of FMR models in market segmentation and the social sciences.

Since there might be unimportant covariates in the pool of variables, and the contribution of each variable to the response might vary among components, variable selection for FMR is of huge importance. Wang et al. (1996) applied Akaike information criterion (AIC) and Bayes information criterion (BIC) in finite mixture of Poisson regression models. Khalili and Chen (2007) introduced a penalized likelihood approach for variable selection in FMR models, and showed the selection method to be consistent. Khalili et al. (2011) studied the problem of feature selection in finite mixture of regression models in large feature spaces, and proposed a 2-stage procedure to overcome the computation complexity.

Model selection is a fundamental part of modern statistics, and so there is a big number of methodologies and extensive literatures on this topic. However, in the presence of outliers or data contamination, it is expected that the classical variable selection methods could be distorted. As a result, \emph{robust} model selection method is becoming increasingly popular. For regression models, some of the approaches focus on modification of selection criteria, such as the Akaike information criterion (Ronchetti, 1985) and Mallows's Cp (Ronchetti and Staudte, 1994), and some on resampling methods (Wisnowski et al., 2003). Ronchetti et al. (1997) proposed robust model selection by cross-validation, and Atkinson and Riani (2002) studied an added-variable t-test for variable selection based on the forward selection. M$\ddot{u}$ller and Welsh (2005) proposed a robust model selection method for linear regression models by using the BIC and bootstrap. By adding a mean shift parameter for each data points, Bondell et al. (2016) studied a methodology that does outlier detection and variable selection simultaneously in linear regression. In addition, Cantoni and Ronchetti (2001) developed robust selection criteria for generalized linear models, and Ronchetti and Trojani (2001) for generalized method of moments. Fan et al. (2012) studied a robust variable selection approach based on a penalized robust estimating equation that incorporates the correlation structure for longitudinal data. Zhang et al. (2013) proposed a robust estimation and variable selection technique for semiparametric partially linear varying coefficient model through modal regression (Yao and Li, 2014).

To the best of our knowledge, however, not much effort has been put on the robust model selection of FMR models. In this article, we develop a robust variable selection method for finite mixture of regression models through the idea of trimmed likelihood estimation. The rest of the article is organized as follows. The derivations of the robust model selection method are given in Section 2. In Section 3, we use simulation studies and real data examples to show the effectiveness of the new methods. A discussion section ends the paper.

\section{New variable selection method for FMR}
\subsection{Penalized FMR}
\label{penalizedfmr}
Let $J$ be a latent class variable with
\begin{equation*}
P(J=j\mid \bx)=\pi_j,
\label{propassump}
\end{equation*}
for $j=1,2,\ldots,m$, where $m$ is the number of components, $\bx$ is
a $(p+1)$-dimensional vector with the first element one, and $\pi_j$ is the proportion of $j$th component such that $\sum_{j=1}^m\pi_j=1$. Given $J=j$, suppose that the
response $y$ depends on $\bx$ in a linear way
$y=\bx^\top\bbeta_j+\epsilon_j,$
$\bbeta_j=(\beta_{0j},\beta_{1j},\ldots,\beta_{pj})^\top$, and
$\epsilon_j\sim N(0,\sigma_j^2)$.  Then the conditional distribution
of $Y$ given $\bx$ without observing $J$ can be written as
\begin{eqnarray}
f(y|\bx,\btheta)=\sum_{j=1}^{m}\pi_{j}\phi(y;\bx^\top\bbeta_{j},\sigma_{j}^{2}),
\label{ch2-eq2-2}
\end{eqnarray}
where $\phi(y;\mu,\sigma^{2})$ denotes density function of $N(\mu,\sigma^2)$, and the log-likelihood function for observations
$\{(\bx_1,y_1),\ldots,(\bx_n,y_n)\}$ is
\begin{equation*}
\ell_n(\btheta)=\sum_{i=1}^n\log\left[\sum_{j=1}^m\pi_j\phi(y_i;\bx_i^\top\bbeta_{j},\sigma_j^2)\right],
\label{mixlinloglh}
\end{equation*}
where $\btheta=(\pi_1,\ldots,\pi_m,\bbeta_1^\top,\ldots,\bbeta_m^\top,\sigma^2_1,\ldots,\sigma^2_m)^\top$.

In order to do variable selection, one commonly used method (Khalili and Chen, 2007) is to maximize the following objective function
\begin{equation}
\ell_1(\btheta)=\ell_n(\btheta)-\bp_n(\btheta),
\label{l1}
\end{equation}
where
\begin{equation*}
\bp_n(\btheta)=\sum_{j=1}^m\sum_{k=1}^p p_{nj}(\beta_{jk}),
\end{equation*}
$p_{nj}(\cdot)$ is some non-negative penalty function. Some of the commonly used penalty functions are:
\begin{itemize}
\item Lasso penalty by Tibshirani (1996): $p_{nj}(\beta)=\lambda_{nj}\sqrt{n}|\beta|$;
\item SCAD penalty by Fan and Li (2001): $p'_{nj}(\beta)=\lambda_{nj}\sqrt{n}I\{\sqrt{n}|\beta|\leq\lambda_{nj}\}+\sqrt{n}(a\lambda_{nj}-\sqrt{n}|\beta|)_+I\{\sqrt{n}|\beta|>\lambda_{nj}\}/(a-1)$;
\item MCP penalty by Zhang (2010): $p'_{nj}(\beta)=\sqrt{n}(\lambda_{nj}-|\beta|/a)I\{\sqrt{n}|\beta|\leq a\lambda_{nj}\}$.
\end{itemize}

By the maximization of $\ell_1(\btheta)$, variable selection and parameter estimation can be done simultaneously, and thus the procedure is computationally efficient. Similar to Fan and Li (2001), in the numerical realization of the technique, we replace $p_{nj}(\beta)$ by a local quadratic approximation
\[\tilde{p}_{nj}(\beta)\approx p_{nj}(\beta_0)+\frac{p'(\beta_0)}{2\beta_0}(\beta^2-\beta_0^2),\]
and the optimization of $\ell_1(\btheta)$ is done through the optimization of
\[Q(\btheta)=\ell_n(\btheta)-\tilde{\bp}_n(\btheta)\]
where $\tilde{\bp}_n(\btheta)=\sum_{j=1}^m\sum_{k=1}^p\tilde{p}_{nj}(\beta_{jk})$, which can be done through an EM algorithm, as follows.


Let $\btheta^{(0)}$ be the initial
value. Starting with $l=0$:\\
\noindent\underline{\textbf{E-step:}}\\
Calculate the expectations of component labels based on estimates from $l^{th}$ iteration:
\begin{equation*}
r_{ij}^{(l+1)}=\frac{\pi_j^{(l)}\phi(y_i;\bx_i^\top\bbeta_j^{(l)},\sigma_j^{2(l)})}{\sum_{j'=1}^m\pi_{j'}^{(l)}\phi(y_i;\bx_i^\top\bbeta_{j'}^{(l)},\sigma_{j'}^{2(l)})}.
\end{equation*}
\underline{\textbf{M-step:}} \\
Update the estimates
\begin{align*}
&\pi_j^{(l+1)}=\frac{\sum_{i=1}^nr_{ij}^{(l+1)}}{n},\\
&\beta_{jk}^{(l+1)}=\arg\max \left[\sum_{i=1}^nr_{ij}^{(l+1)}\log\phi(y_i;\bx_i^\top\bbeta_j,\sigma_j^{2(l)})-\sum_{k=1}^p\tilde{p}_{nj}(\beta_{jk})\right],\\
&\sigma_j^{2(l+1)}=\frac{\sum_{i=1}^n(y_i-\bx_i^\top\bbeta_j^{(l+1)})^2r^{(l+1)}_{ij}}{\sum_{i=1}^n r^{(l+1)}_{ij}},
\end{align*}
for $j=1,\ldots,m$. The following theorem proves the monotonicity of the above algorithm with the proof provided in the Appendix.

\begin{thm}
\label{thm1}
Suppose that $p_{nj}$ on $(0,\infty)$ is piecewise differentiable, nondecreasing and concave. Furthermore, $p_{nj}$ is continuous at 0 and $p_{nj}'(0+)<\infty$.
Then, the objective function (\ref{l1}) is non-decreasing after each iteration of the algorithm, i.e.,
\[\ell_1(\btheta^{(l+1)})\geq\ell_1(\btheta^{(l)}),\]
until a fixed point is reached.
\end{thm}

\subsection{Robust model selection through trimming}
The maximum likelihood estimator (MLE) via the expectation maximization (EM) algorithm is the most commonly used method for finite mixture of regression models, but it is sensitive to outliers. Neykov et al. (2007) proposed the trimmed likelihood estimator (TLE) of mixture models, which only uses $(1-\alpha)\times100\%$ of the data to fit the model, and removes the remaining $\alpha\times100\%$ observations that are highly unlikely to occur if the fitted model were true. Li et al. (2016) investigated a robust estimation of the number of components in the mixture of regression models using trimmed information criteria. Yang et al. (2017) introduced a robust estimation procedure of mixtures of factor analyzers using the trimmed likelihood estimator.

Applying the idea of trimmed likelihood estimator, we propose a robust model selection method through trimming, which can be expressed as:
\begin{equation*}
\max_{I\in I_\alpha}\max_{\btheta}\sum_{i\in I}\log f(y_i|\bx_i,\btheta)-\tilde{\bp}_n(\btheta),
\end{equation*}
where $f(y|\bx,\theta)$ is the density defined in (\ref{ch2-eq2-2}), and $I_\alpha$ is the set of all $\left \lfloor n(1-\alpha) \right \rfloor$-subsets of the set $(1,\ldots,n)$.

To overcome the combinatorial nature of TLE, that is, all possible $\binom{n}{\left \lfloor n(1-\alpha) \right \rfloor}$ combinations have to be fitted, we extend the FAST-TLE algorithm (M$\ddot{u}$ller and Neykov 2003; Neykov et al., 2007) to the penalized mixture regression setting for an approximation. The detailed algorithm is summarized as follows:

\begin{table}[htb]
    \centering
\def\arraystretch{1.5}
\small \hspace*{-22.75pt}
\begin{tabular}{l }
\hline
{\bf Algorithm 1:}\\
\hline
Find an initial value of $\btheta$, denoted by $\btheta_0$. \\
Sort $f(y|\bx,\btheta_0)$ as $f(y_{\nu(1)}|\bx_{\nu(1)},\btheta_0)\geq\ldots\geq f(y_{\nu(n)}|\bx_{\nu(n)},\btheta_0)$, then $\{\nu(1),\ldots,\nu(\left \lfloor n(1-\alpha) \right \rfloor)\}$ forms the \\
\hskip 0.2in index set $\hat{I}_\alpha$.\\
{\bf while} change in estimators $\geq $ c {\bf do}\\
\hskip 0.2in Given an index set $\hat{I}_\alpha$, apply the EM algorithm introduced in Section \ref{penalizedfmr} to $\tilde{\bx}=\bx[\hat{I}_\alpha]$, $\tilde{y}=y[\hat{I}_\alpha]$ to update\\
\hskip 0.4in the estimator $\hat{\btheta}$, where $\tilde{\bx}$ and $\tilde{y}$ mean the data matrix only containing the rows indexed by $\hat{I}_\alpha$.\\
\hskip 0.2in For an estimator $\hat{\btheta}$, sort $f(y|\bx,\hat{\btheta})$ as $f(y_{\nu(1)}|\bx_{\nu(1)},\hat{\btheta})\geq\ldots\geq f(y_{\nu(n)}|\bx_{\nu(n)},\hat{\btheta})$,\\
\hskip 0.4in then $\{\nu(1),\ldots,\nu(\left \lfloor n(1-\alpha) \right \rfloor)\}$ forms the index set $\hat{I}_\alpha$.\\
{\bf end while}\\
\hline
\end{tabular}
\end{table}

\section{Numerical studies}
\subsection{Simulation study}

In this section, we use a simulation study to show the finite sample performance of the proposed robust model selection through trimming, and compare it with the regular variable selection method for FMR model. We consider two-component FMR models, where $\pi_1=0.5$ or $0.7$, $\bbeta$'s are listed in Table \ref{models}, and $\sigma_1^2=\sigma_2^2=1$. $\bx$ is generated from a multivariate normal with mean 0, variance 1, and correlation structures either $\rho_{ij}=\text{cor}(x_i,x_j)=0.5^{|i-j|}$ or $\rho_{ij}=0$.

\begin{table}[htb]
    \centering
\caption{Model setting.}\vskip 0.05in
\def\arraystretch{1.5}
\small \hspace*{-22.75pt}
\begin{tabular}{c c c } \hline
parameters & Model 1 & Model 2 \\
\hline
$\beta_1$ & (1, 0, 0, 3, 0) & (1, 0.6, 0, 3, 0) \\
$\beta_2$ & (-1, 2, 0, 0, 3) & (-1, 0, 0, 4, 0.7) \\
\hline
\end{tabular}
\label{models}
\end{table}

For simulation studies, a total of 200 random samples with sample sizes of $n=100$ and $n=200$ are generated for each case. To show the robustness of our method, the following three contamination schemes are considered. Namely,

\begin{itemize}
\item Contamination 1-3: perturb the $\alpha_0*n$ responses by adding a random number from $U(7,10)$ to the original responses, where $\alpha_0=0.01,0.03$ and $0.05$, respectively.
\end{itemize}

To measure the performance of model selection, the average of number of correct zeros and incorrect zeros of regression coefficients are reported. In addition, the median of model error (MME), defined by $(\hat{\bbeta}-\bbeta_0)^\top {\rm E}(XX^\top)(\hat{\bbeta}-\bbeta_0)$, and model accuracy, defined by the proportion of times when the exact model is selected, are also reportsed. $\alpha=0.05$ is applied throughout the study, and the estimation results are summarized in Table \ref{Table1}-\ref{Table8}. 
It can be seen that, the new method has better model selection performance and provides smaller model errors across all modeling settings. 


\subsection{Selection of trimming proportion}

\begin{figure}[htbp]
    \centering
    \includegraphics[width=1\textwidth,height=0.4\textheight]{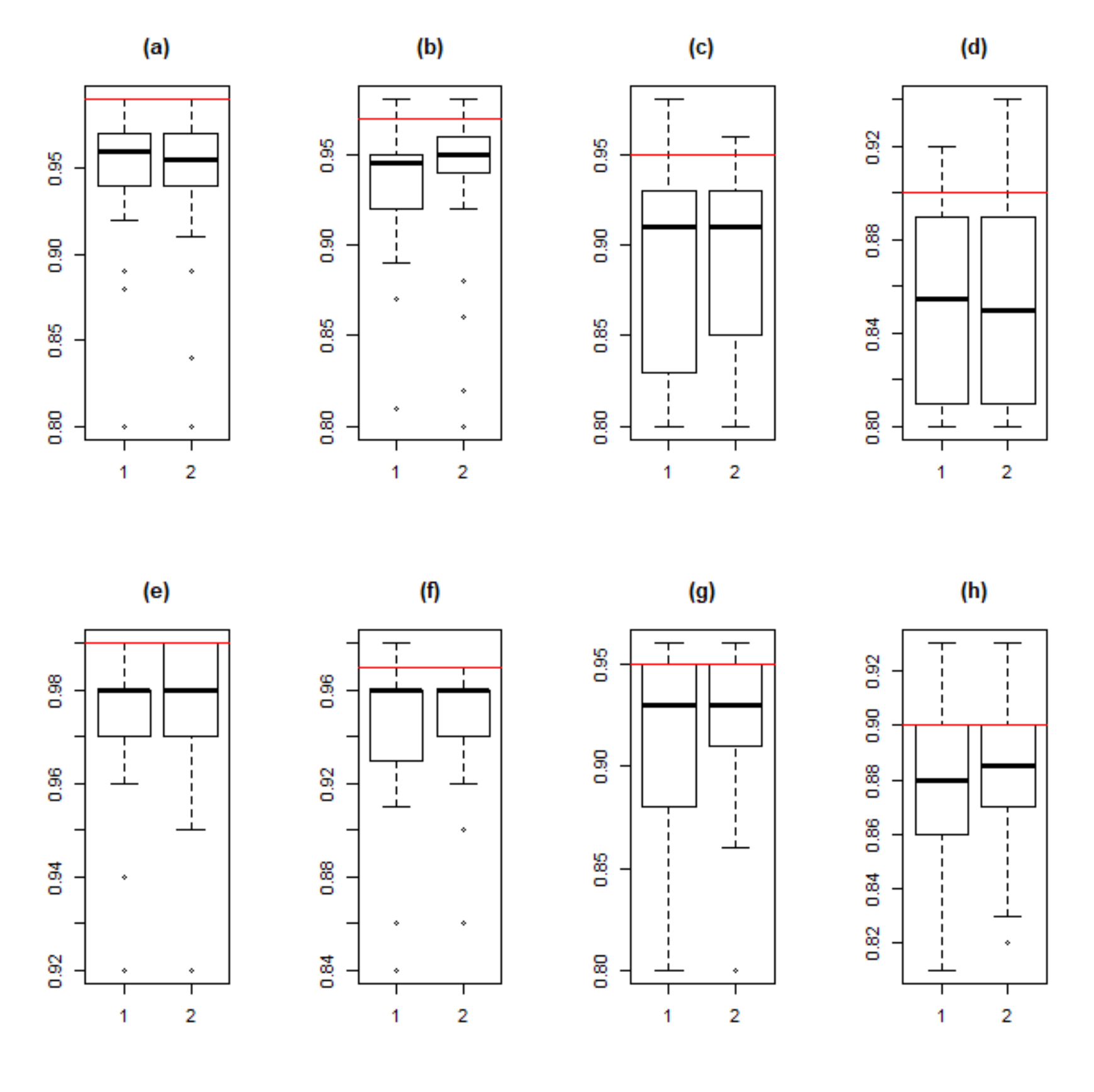}
   \label{selalpha}
  \caption{Boxplot of ``1-$\alpha$'' selected by minimizing the maximum of (1) diagonal values; (2) eigenvalues: (a) $\alpha_0=0.01, n=100$; (b) $\alpha_0=0.03, n=100$; (c) $\alpha_0=0.05, n=100$; (d) $\alpha_0=0.1, n=100$; (e) $\alpha_0=0.01, n=200$; (f) $\alpha_0=0.03, n=200$; (g) $\alpha_0=0.05, n=200$; (h) $\alpha_0=0.1, n=200$. The red line indicates ``1-$\alpha_0$''.}
\end{figure}

In the previous study, the trimming proportion $\alpha$ is assumed to be known and fixed. In reality, however, the selection of $\alpha$ is a difficult task. Neykov et al. (2007) and Li et al. (2016) applied a graphical tool. After making a curve of trimmed likelihood versus $\alpha$'s, they proposed to estimate the trimming proportion by the largest $\alpha$ at which the slope of the curve changes. However, as Neykov et al. (2007) pointed out, this method tends to underestimate the true values in some cases. In this article, we proposed a bootstrap procedure to choose the trimming proportion by minimizing the variability of the estimators. A brief description of the bootstrap procedure is summarized in Algorithm 2. The result of a simulation study is summarized in Figure \ref{selalpha} and Table \ref{selalphat}, where 200 bootstrap samples were estimated at a grid points of $\alpha$ values, ranging from 0 to 20\% in steps of 1\% over 30 repetitions.

\begin{table}[htb]
    \centering
\def\arraystretch{1.5}
\small \hspace*{-22.75pt}
\begin{tabular}{l }
\hline
{\bf Algorithm 2:}\\
\hline
Generate a set of data with size $n$, say $(x,y)$. \\
Generate a grid of alpha's, say $alpha.grid=(0.5,0.01,0.99)$.\\
{\bf for} $alpha$ in $alpha.grid$\\
\hskip 0.2in {\bf for} t.boot=1:time.boot\\
\hskip 0.4in Sample $n$ observations with replacement from $(x,y)$, and call the generated dataset $(x.boot,y.boot)$. \\
\hskip 0.4in Apply Algorithm 1 to $(x.boot,y.boot)$ with alpha, and name the estimators (bet.boot,pr.boot,sig.boot).\\
\hskip 0.2in {\bf end for}\\
\hskip 0.2in Calculate the covariance matrices of each component of bet.boot, say $(cov_1,\ldots,cov_m)$.\\
{\bf end for}\\
Find $alpha$, which minimizes the maximum of diagonal values or eigenvalues of $(var_1,\ldots,var_m)$.\\
\hline
\end{tabular}
\end{table}

\begin{table}[htb]
\renewcommand{\arraystretch}{1.5}
\caption{\small  Alpha's selected by cross-validation.} \label{selalphat}
\begin{center}
\scriptsize
\setlength{\tabcolsep}{1mm}
\begin{tabular}{c  cc| cc|cc| cc}
\hline
 &\multicolumn{2}{c}{$\alpha_0=0.01$ } &\multicolumn{2}{c}{$\alpha_0=0.03$} &\multicolumn{2}{c}{$\alpha_0=0.05$} &\multicolumn{2}{c}{$\alpha_0=0.1$} \\
\cline{2-3}\cline{4-5}\cline{6-7}\cline{8-9}
  &median & mean &median & mean &median & mean &median & mean \\
\hline
maximum of diagonal values\\
$n=100$& 0.960&0.951&0.945&0.931&0.910&0.889&0.855&0.851\\
$n=200$&0.980&0.974&0.960&0.943&0.930&0.915&0.880&0.875\\
maximum of eigen values\\
$n=100$&0.955&0.945&0.955&0.934&0.910&0.893&0.850&0.853\\
$n=200$&0.980&0976&0.960&0.945&0.930&0.920&0.995&0.852\\
\hline
\end{tabular}
\end{center}
\end{table}

\subsection{Real data applications}

\emph{Example 1 (Baseball salary data)}. We apply our methodology to the baseball salary data, used by Khalili and Chen (2007) and Jiang (2016), which is available at www. amstat. org/publications/jse. The dataset contains the salaries (thousands of dollars) of $n=337$ major league baseball players who played at least one game in both the 1991 and 1992 seasons, excluding pitchers. Along with it are 16 performance measures: batting average ($X_1$), on-base percentage ($X_2$), runs ($X_3$), hits ($X_4$), doubles ($X_5$), triples ($X_6$), home runs ($X_7$), runs batted in ($X_8$), walks ($X_9$), strikeouts ($X_{10}$), stolen bases($X_{11}$) , errors ($X_{12}$), and indicators of free agency eligibility ($X_{13}$), free agent in 1991/2 ($X_{14}$), arbitration eligibility ($X_{15}$), and arbitration in 1991/2 ($X_{16}$). Following Khalili and Chen (2007), we also consider interactions of the four indicators $X_{13}-X_{16}$ and the quantitative variables $X_1, X_3, X_7$, and $X_8$, which leads to a total of 32 potential covariates that affect players' salary.

Since the distribution of the original response is highly right-skewed, a log transformation (actually $\log(x+0.1)$) is applied to the salary to give a new response. Similar to Khalili and Chen (2007), as suggested by the histogram of the transformed response, a two-component mixture of regressions model is assumed for the data, and model selection results of mixture model with lasso and scad penalty (denoted by ``ml'' and ``ms'', respectively), and the newly proposed robust method (denoted by ``mtl'' and ``mts'') are applied to select predictors. The parameter estimators of different methods are listed in Table \ref{salaryest}, and the mean squared prediction errors based on 10-fold cross-validation and Monte Carlo cross-validation (MCCV, Shao, 1993) with $d=100$ are shown in Figure 1 (a) and (d). It can be seen that the robust variable selection methods work comparably to, though slightly worse than, the existing methods.

To compare the robustness of different methods, similar to Li et al. (2016), 1\% or 5\% of random noise from $U(7,10)$ are added to the original response. First, we apply the bootstrap method proposed above to select the trimming proportions, and $\alpha=0.03$ and $\alpha=0.09$ are selected, which is slightly conservative. The mean squared prediction errors are shown in Figure 1. Clearly, when outliers are presented in the dataset, the robust variable selection methods perform much better than the existing methods.

\begin{table}[htb]
\renewcommand{\arraystretch}{1.5}
\caption{\small  Baseball salary data: parameter estimators.} \label{salaryest}
\begin{center}
\scriptsize
\setlength{\tabcolsep}{1mm}
\begin{tabular}{c  cc| cc|cc| cc}
\hline
 &\multicolumn{2}{c}{ml } &\multicolumn{2}{c}{ms} &\multicolumn{2}{c}{mtl } &\multicolumn{2}{c}{mts} \\
\cline{2-3}\cline{4-5}\cline{6-7}\cline{8-9}
 Covariate  &Comp 1&Comp 2 &Comp 1&Comp 2 &Comp 1&Comp 2 &Comp 1&Comp 2 \\
\hline
$X_0$ & 7.304 & 5.097 & 6.091 & 4.881 & 5.067 & 4.827 & 5.603 & 4.769\\
$X_1$ & & & & & -0.667 & & 6.210 & -0.305\\
$X_2$ & & & & -0.487 & 3.479 & -0.269 & -3.902 & \\
$X_3$ & 0.006 & -0.005 & 0.003 & -0.003 & & & & \\
$X_4$ & -0.003 & 0.007 & & 0.009 & & & & \\
$X_5$ & 0.009 & -0.007 & & -0.010 & & & & \\
$X_6$ & -0.016 & & & & & & & \\
$X_7$ & -0.012 & & & & & & & \\
$X_8$ & 0.023 & 0.005 & 0.010 & 0.001 & & & & \\
$X_9$ & -0.003 & 0.004 & & 0.004 & & & & \\
$X_{10}$ & -0.036 & 0.001 & -0.011 & -0.001 & & & & \\
$X_{11}$ & -0.001 & 0.003 & & 0.001 & & & & \\
$X_{12}$ & & -0.007 & & & & & & \\
$X_{13}$ & & 1.027 & & 2.987 & 0.839 & 2.696 & 1.479 & 2.471\\
$X_{14}$ & & & & -3.917 & & -3.351 & -0.359 & -2.897\\
$X_{15}$ & 0.593 & 3.151 & 1.094 & 5.582 & 1.146 & 3.876 & 1.087\\
$X_{16}$ & & & -6.300 & 1.187 & -3.020 & 3.851 & -4.836 & 5.082\\
$X_1*X_{13}$ & & & & -3.378 & -3.084 & -1.369 & -4.567 & 1.542 \\
$X_3*X_{13}$ & 0.007 & 0.007 & 0.002 & & & & & \\
$X_7*X_{13}$ & & 0.012 & & 0.006 & & & & \\
$X_8*X_{13}$ & 0.004 & 0.004 & 0.014 & & & & & \\
$X_1*X_{14}$ & & & & 11.162 & 12.355 & 7.548 & 6.642 & 3.109\\
$X_3*X_{14}$ & 0.001 & 0.003 & -0.001 & 0.010 & & & & \\
$X_7*X_{14}$ & & 0.039 & & & & & & \\
$X_8*X_{14}$ & & -0.012 & & 0.003 & & & & \\
$X_1*X_{15}$ & & & -8.025 & & -16.071 & & -8.246 & 0.575\\
$X_3*X_{15}$ & & 0.006 & & & & & & \\
$X_7*X_{15}$ & & -0.003 & & & & & & \\
$X_8*X_{15}$ & -0.006 & 0.011 & & 0.010 & & & & \\
$X_1*X_{16}$ & &  & 23.043 & -7.319 & 11.356 & -18.420 & 17.196 & -23.488\\
$X_3*X_{16}$ & 0.018 & -0.005 & & 0.003 & & & & \\
$X_7*X_{16}$ & 0.012 & & & & & & & \\
$X_8*X_{16}$ & & 0.003 & & 0.009 & & & & \\
\hline
\end{tabular}
\end{center}
\end{table}

\begin{figure}[htbp]
    \centering
    \includegraphics[width=1\textwidth,height=0.4\textheight]{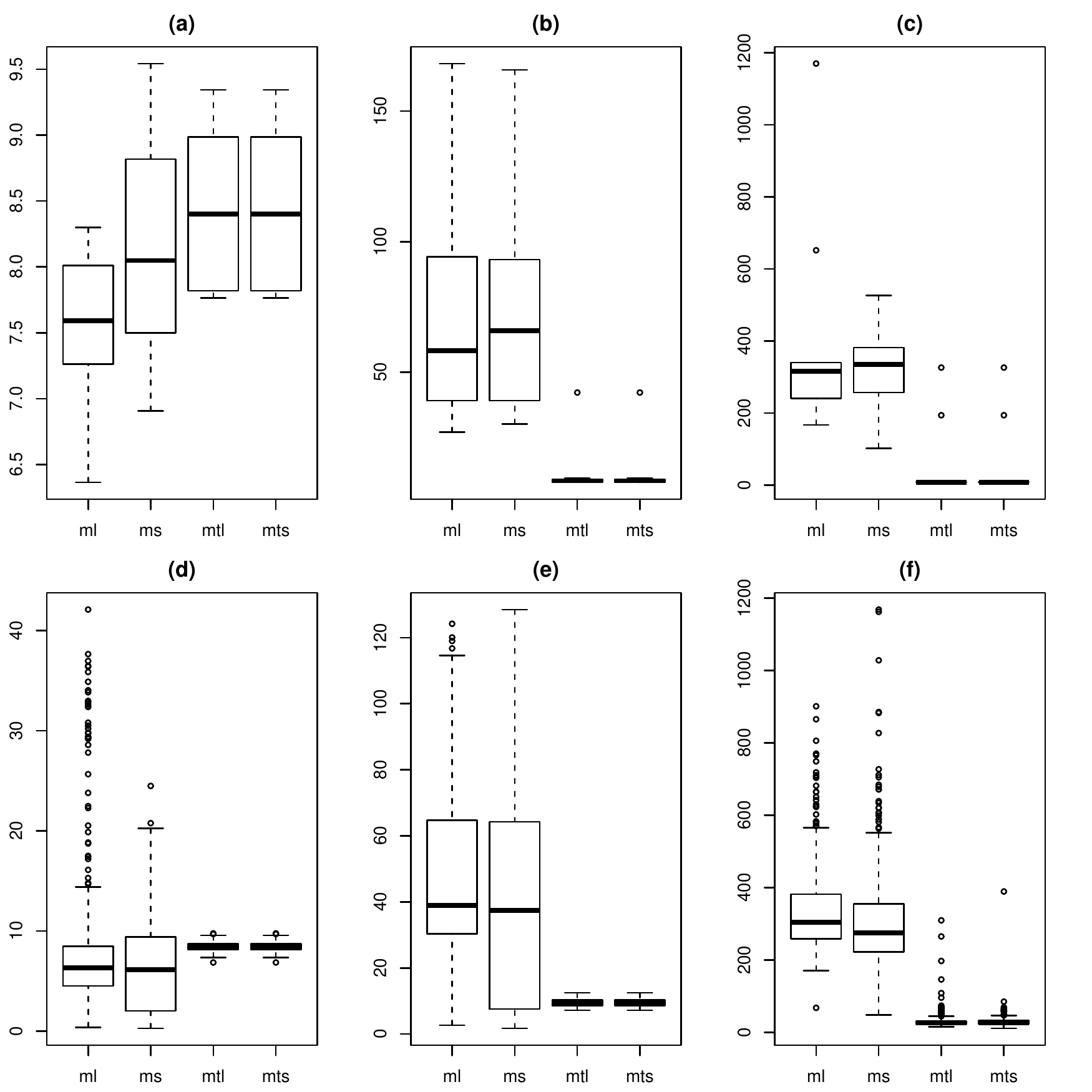}
   \label{noncont}
  \caption{Baseball salary data: mean squared prediction error by (a) 10-fold CV (no contamination); (b) 10-fold CV (1\% contamination); (c) 10-fold CV (5\% contamination); (d) MCCV with $d=100$ (no contamination); (e) MCCV with $d=100$ (1\% contamination); (f) MCCV with $d=100$ (5\% contamination).}
\end{figure}

\emph{Example 2 (Hong Kong air pollution data)}. Next, we apply the new methods to the Hong Kong air pollution data, which has been analyzed by Fan and Zhang (1999), Cai et al. (2000), Xia et al. (2002), among others. The dataset contains the daily air pollutants and other environmental facts of Hong Kong from January 1, 1994 to December 31, 1997 (and so $n=1461$). Of interest is how air conditions and other factors affect daily hospital admissions for respiratory diseases. The potential covariates are : daily admissions for cardiovascular disease ($X_1$), concentration of NO2 ($X_2$), concentration of SO2 ($X_3$), daily admissions for circulatory diseases ($X_4$), concentration of O3 ($X_5$), temperature ($X_6$), humidity ($X_7$), the day of the month ($X_8$), and the month of the year ($X_9$). Since skewness is shown in the histograms of the response and covariates $X_1$-$X_7$, the analysis is done to the log scale of those variables.

The robust model selection method proposed by Li et al. (2016) indicates that a two-component FMR model is appropriate for the data. The bootstrap method shows that 9\% should be used to trim the data, and the parameter estimates and prediction performance based on the original dataset are shown in Table \ref{hkest} and Figure 2. As pointed out by Yang et al. (2017), 166 observations (166/1461=11.3\%) are detected as outliers, and so it is no surprise that the robust variable selection methods outperform the existing ones.

\begin{table}[htb]
\renewcommand{\arraystretch}{1.5}
\caption{\small  Hong Kong air pollution data: parameter estimators.} \label{hkest}
\begin{center}
\scriptsize
\setlength{\tabcolsep}{1mm}
\begin{tabular}{c  cc| cc|cc| cc}
\hline
 &\multicolumn{2}{c}{ml } &\multicolumn{2}{c}{ms} &\multicolumn{2}{c}{mtl} &\multicolumn{2}{c}{mts} \\
\cline{2-3}\cline{4-5}\cline{6-7}\cline{8-9}
 Covariate  &Comp 1&Comp 2 &Comp 1&Comp 2 &Comp 1&Comp 2 &Comp 1&Comp 2 \\
\hline
$X_0$ & 3.053 & 5.594 & 1.541 & 1.231 & 0.723 & 1.449 & 0.723 & 1.449\\
$X_1$ & 0.444 & & 0.718 & 0.682 & 0.700 & 0.686 & 0.700 & 0.686\\
$X_2$ & & & & & & 0.151 & & 0.151\\
$X_3$ & & & & & -0.054 & & -0.054\\
$X_4$ & & & & & & & & -0.041\\
$X_5$ & & & & & & & & \\
$X_6$ & & & & & 0.091 & 0.112 & 0.091 & 0.112\\
$X_7$ & & & & 0.169 & 0.249 & & 0.249 & -0.042\\
$X_8$ & & -0.013 & & & -0.027 & & & \\
$X_9$ & & & & & & & & \\
\hline
\end{tabular}
\end{center}
\end{table}

\begin{figure}[htbp]
    \centering
    \includegraphics[width=0.5\textwidth,height=0.4\textheight]{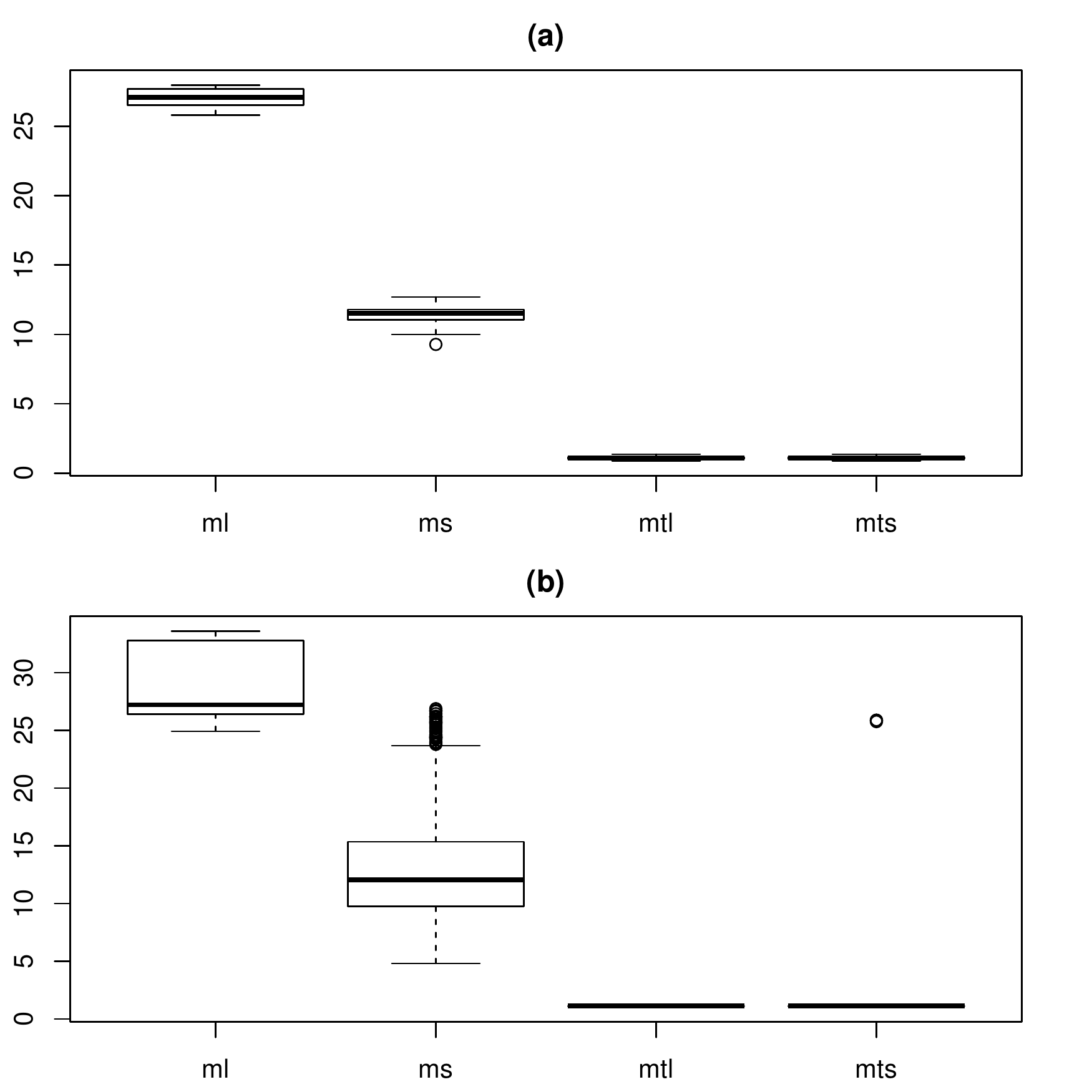}
   \label{Figure1}
  \caption{Hong Kong air pollution data: mean squared prediction error by (a) 10-fold CV and (b) MCCV with $d=300$. }
\end{figure}

\section{Discussion}
\label{sec:discussion}
The method proposed in this article combines the ideas of trimming and penalized finite mixture of regressions model and enables us to perform model selection for FMR model robustly. We demonstrated the superiority of the trimmed methods in comparison with the traditional one when data are contaminated using a simulation study and two real data examples.

In this article, we applied a pre-chosen trimming proportion $\alpha$ in our numerical studies. As pointed out by Neykov et al. (2007) and Li et al. (2016), graphical tools could be used to choose $\alpha$, but it tends to underestimate the true values in some cases. Therefore, further work could be done on how to adaptively choose an optimal or conservative trimming proportion $\alpha$.

It would also be interesting to see if the trimming idea could be applied to do model selection and parameter estimation for some more complicated semiparametric mixture of regression models, such as the mixture of regression models with varying proportions (Huang and Yao, 2012), mixture of regressions model with nonparametric errors (Hunter and Young, 2012), FMR models with single-index (Xiang and Yao, 2017), and so on.

\section*{Appendix A}
The following proof of Theorem \ref{thm1} indicates that the objective function (\ref{l1}) is nondecreasing in each iteration of the algorithm.

\textbf{Proof of Theorem \ref{thm1}}: Let $\ell_c(\btheta|\btheta^{(l)})=\sum_{i=1}^n\sum_{j=1}^mr_{ij}^{(l+1)}\log\pi_j\phi(y_i;\bx_i^\top\bbeta_j,\sigma_j^2)$ be the complete log-likelihood conditional on the current estimates $\btheta^{(l)}$. Then, note that
\begin{align*}
\ell_c(\btheta|\btheta^{(l)})=&\sum_{i=1}^n\sum_{j=1}^mr_{ij}^{(l+1)}\log\pi_j\phi(y_i;\bx_i^\top\bbeta_j,\sigma_j^2)\\
\leq & \sum_{i=1}^n\sum_{j=1}^m\log\pi_j\phi(y_i;\bx_i^\top\bbeta_j,\sigma_j^2)\\
\leq & \sum_{i=1}^n\log\sum_{j=1}^m\pi_j\phi(y_i;\bx_i^\top\bbeta_j,\sigma_j^2)=\ell_n(\btheta).
\end{align*}

Then, $\ell_c(\btheta|\btheta^{(l)})-\tilde{\bp}(\btheta)$, the objective function maximized in the M step, minorizes $Q(\btheta)$ at $\btheta^{(l)}$. In addition, by Corollary 3.1 of Hunter and Li (2005), if all $p_{nk}$'s satisfy the conditions given in the theorem, $Q(\btheta)$  minorizes $\ell_1(\btheta)$ at $\btheta^{(l)}$. Therefore, by the property of MM algorithm, $\ell_1(\btheta^{(l+1)})\geq \ell_1(\btheta^{(l)})$.

\section*{Appendix B}
The following tables summarizes the simulation results.

\begin{sidewaystable}[htb]
\renewcommand{\arraystretch}{1.5}
\caption{\small  Variable selection results for Model 1 with $n=100_{(n=200)}$, $\pi_1=0.5$, and $\rho_{ij}=0$.} \label{Table1}
\begin{center}
\scriptsize
\setlength{\tabcolsep}{1mm}
\begin{tabular}{c  cc| cc|cc| cc|   cc| cc|cc| cc}
\hline\hline
        &\multicolumn{8}{c}{FMR}      &\multicolumn{8}{|c}{trim}  \\
\cline{2-9}\cline{10-17}
 &\multicolumn{2}{c}{Correct } &\multicolumn{2}{c}{Incorrect} &\multicolumn{2}{c}{MME } &\multicolumn{2}{c}{Model Accuracy} &\multicolumn{2}{|c}{Correct } &\multicolumn{2}{c}{Incorrect}    &\multicolumn{2}{c}{MME } &\multicolumn{2}{c}{Model Accuracy}    \\
\cline{2-3}\cline{4-5}\cline{6-7}\cline{8-9}\cline{10-11}\cline{12-13}\cline{14-15}\cline{16-17}
 Method  &Comp1(3)&Comp2(2)&Comp1&Comp2 &Comp1&Comp2 &Comp1&Comp2 &Comp1(3)&Comp2(2)&Comp1&Comp2 &Comp1&Comp2 &Comp1&Comp2\\
\hline
 Cont 1\\
 LASSO & 2.58$_{(2.87)}$ & 1.76$_{(1.94)}$ & 0.01$_{(0.00)}$ & 1.00$_{(1.00)}$ & 0.05$_{(0.01)}$ & 0.09$_{(0.03)}$ & 0.67$_{(0.88)}$ & 0.01$_{(0.00)}$ & 2.50$_{(2.89)}$ & 1.72$_{(1.89)}$ & 0.00$_{(0.00)}$ & 0.00$_{(0.00)}$ & 0.06$_{(0.01)}$ & 0.09$_{(0.03)}$ & 0.61$_{(0.90)}$ & 0.75$_{(0.89)}$\\
 SCAD & 2.55$_{(2.84)}$ & 1.73$_{(1.92)}$ & 0.01$_{(0.00)}$ & 1.00$_{(1.00)}$ & 0.06$_{(0.01)}$ & 0.09$_{(0.03)}$ & 0.65$_{(0.86)}$ & 0.01$_{(0.00)}$ & 2.40$_{(2.82)}$ & 1.65$_{(1.87)}$ & 0.00$_{(0.00)}$ & 0.00$_{(0.00)}$ & 0.08$_{(0.01)}$ & 0.10$_{(0.03)}$ & 0.55$_{(0.84)}$ & 0.70$_{(0.87)}$\\
 MCP & 2.55$_{(2.85)}$ & 1.73$_{(1.92)}$ & 0.01$_{(0.00)}$ & 1.00$_{(1.00)}$ & 0.06$_{(0.01)}$ & 0.10$_{(0.03)}$ & 0.60$_{(0.86)}$ & 0.01$_{(0.00)}$ & 2.40$_{(2.82)}$ & 1.65$_{(1.87)}$ & 0.00$_{(0.00)}$ & 0.00$_{(0.00)}$ & 0.08$_{(0.01)}$ & 0.10$_{(0.03)}$ & 0.55$_{(0.84)}$ & 0.70$_{(0.87)}$\\
 \hline
 Cont 2\\
 LASSO & 2.27$_{(2.40)}$ & 1.75$_{(1.88)}$ & 0.04$_{(0.01)}$ & 0.98$_{(1.00)}$ & 0.15$_{(0.10)}$ & 0.12$_{(0.04)}$ & 0.47$_{(0.54)}$ & 0.01$_{(0.00)}$ & 2.49$_{(2.84)}$ & 1.65$_{(1.88)}$ & 0.00$_{(0.00)}$ & 0.00$_{(0.00)}$ & 0.04$_{(0.01)}$ & 0.09$_{(0.03)}$ & 0.60$_{(0.85)}$ & 0.69$_{(0.89)}$\\
 SCAD & 2.24$_{(2.37)}$ & 1.71$_{(1.87)}$ & 0.03$_{(0.01)}$ & 0.98$_{(1.00)}$ & 0.15$_{(0.11)}$ & 0.12$_{(0.04)}$ & 0.44$_{(0.51)}$ & 0.01$_{(0.00)}$ & 2.38$_{(2.79)}$ & 1.62$_{(1.85)}$ & 0.00$_{(0.00)}$ & 0.00$_{(0.00)}$ & 0.08$_{(0.01)}$ & 0.10$_{(0.03)}$ & 0.52$_{(0.81)}$ & 0.66$_{(0.86)}$\\
 MCP & 2.24$_{(2.36)}$ & 1.71$_{(1.87)}$ & 0.03$_{(0.01)}$ & 0.98$_{(1.00)}$ & 0.15$_{(0.11)}$ & 0.12$_{(0.04)}$ & 0.44$_{(0.51)}$ & 0.01$_{(0.00)}$ & 2.38$_{(2.79)}$ & 1.62$_{(1.85)}$ & 0.00$_{(0.00)}$ & 0.00$_{(0.00)}$ & 0.08$_{(0.01)}$ & 0.10$_{(0.03)}$ & 0.52$_{(0.81)}$ & 0.66$_{(0.86)}$\\
\hline
 Cont 3\\
 LASSO & 2.08$_{(2.21)}$ & 1.57$_{(1.79)}$ & 0.08$_{(0.04)}$ & 0.91$_{(0.97)}$ & 0.33$_{(0.23)}$ & 0.21$_{(0.04)}$ & 0.38$_{(0.43)}$ & 0.03$_{(0.01)}$ & 2.40$_{(2.91)}$ & 1.58$_{(1.89)}$ & 0.01$_{(0.00)}$ & 0.00$_{(0.00)}$ & 0.07$_{(0.01)}$ & 0.08$_{(0.02)}$ & 0.57$_{(0.92)}$ & 0.67$_{(0.90)}$\\
 SCAD &2.07$_{(2.18)}$ & 1.55$_{(1.77)}$ & 0.08$_{(0.03)}$ & 0.90$_{(0.97)}$ & 0.32$_{(0.24)}$ & 0.22$_{(0.05)}$ & 0.38$_{(0.42)}$ & 0.03$_{(0.01)}$ & 2.27$_{(2.84)}$ & 1.55$_{(1.88)}$ & 0.00$_{(0.00)}$ & 0.00$_{(0.00)}$ & 0.08$_{(0.01)}$ & 0.08$_{(0.02)}$ & 0.50$_{(0.87)}$ & 0.64$_{(0.89)}$\\
 MCP & 2.07$_{(2.18)}$ & 1.55$_{(1.77)}$ & 0.08$_{(0.03)}$ & 0.90$_{(0.97)}$ & 0.32$_{(0.24)}$ & 0.22$_{(0.05)}$ & 0.38$_{(0.42)}$ & 0.03$_{(0.01)}$ & 2.27$_{(2.85)}$ & 1.55$_{(1.87)}$ & 0.00$_{(0.00)}$ & 0.00$_{(0.00)}$ & 0.08$_{(0.01)}$ & 0.08$_{(0.02)}$ & 0.50$_{(0.87)}$ & 0.64$_{(0.89)}$\\
 \hline
\end{tabular}
\end{center}
\end{sidewaystable}

\begin{sidewaystable}[htb]
\renewcommand{\arraystretch}{1.5}
\caption{\small  Variable selection results for Model 1 with $n=100_{(n=200)}$, $\pi_1=0.7$, and $\rho_{ij}=0$.} \label{Table2}
\begin{center}
\scriptsize
\setlength{\tabcolsep}{1mm}
\begin{tabular}{c  cc| cc|cc| cc|   cc| cc|cc| cc}
\hline\hline
        &\multicolumn{8}{c}{FMR}      &\multicolumn{8}{|c}{trim}  \\
\cline{2-9}\cline{10-17}
 &\multicolumn{2}{c}{Correct } &\multicolumn{2}{c}{Incorrect} &\multicolumn{2}{c}{MME } &\multicolumn{2}{c}{Model Accuracy} &\multicolumn{2}{|c}{Correct } &\multicolumn{2}{c}{Incorrect}    &\multicolumn{2}{c}{MME } &\multicolumn{2}{c}{Model Accuracy}    \\
\cline{2-3}\cline{4-5}\cline{6-7}\cline{8-9}\cline{10-11}\cline{12-13}\cline{14-15}\cline{16-17}
 Method  &Comp1(3)&Comp2(2)&Comp1&Comp2 &Comp1&Comp2 &Comp1&Comp2 &Comp1(3)&Comp2(2)&Comp1&Comp2 &Comp1&Comp2 &Comp1&Comp2\\
\hline
 Cont 1\\
 LASSO & 2.78$_{(2.96)}$ & 1.51$_{(1.75)}$ & 0.00$_{(0.01)}$ & 1.00$_{(0.99)}$ & 0.02$_{(0.01)}$ & 0.23$_{(0.07)}$ & 0.81$_{(0.95)}$ & 0.01$_{(0.00)}$ & 2.78$_{(2.97)}$ & 1.29$_{(1.63)}$ & 0.00$_{(0.00)}$ & 0.01$_{(0.00)}$ & 0.02$_{(0.01)}$ & 0.28$_{(0.08)}$ & 0.79$_{(0.97)}$ & 0.42$_{(0.67)}$\\
 SCAD & 2.75$_{(2.93)}$ & 1.49$_{(1.72)}$ & 0.00$_{(0.01)}$ & 1.01$_{(0.99)}$ & 0.02$_{(0.01)}$ & 0.24$_{(0.08)}$ & 0.79$_{(0.94)}$ & 0.01$_{(0)}$ & 2.68$_{(2.95)}$ & 1.18$_{(1.58)}$ & 0.00$_{(0.00)}$ & 0.00$_{(0.00)}$ & 0.02$_{(0.01)}$ & 0.31$_{(0.09)}$ & 0.72$_{(0.95)}$ & 0.37$_{(0.63)}$\\
 MCP & 2.75$_{(2.93)}$ & 1.49$_{(1.72)}$ & 0.00$_{(0.01)}$ & 1.01$_{(0.99)}$ & 0.02$_{(0.01)}$ & 0.24$_{(0.08)}$ & 0.79$_{(0.94)}$ & 0.01$_{(0.00)}$ & 2.68$_{(2.95)}$ & 1.18$_{(1.57)}$ & 0.00$_{(0.00)}$ & 0.00$_{(0.00)}$ & 0.02$_{(0.01)}$ & 0.31$_{(0.09)}$ & 0.72$_{(0.95)}$ & 0.37$_{(0.63)}$\\
\hline
 Cont 2\\
 LASSO & 2.58$_{(2.76)}$ & 1.44$_{(1.49)}$ & 0.08$_{(0.02)}$ & 0.83$_{(0.77)}$ & 0.04$_{(0.01)}$ & 0.68$_{(0.29)}$ & 0.66$_{(0.82)}$ & 0.05$_{(0.07)}$ & 2.61$_{(2.98)}$ & 1.05$_{(1.68)}$ & 0.00$_{(0.00)}$ & 0.03$_{(0.00)}$ & 0.02$_{(0.00)}$ & 0.33$_{(0.08)}$ & 0.72$_{(0.98)}$ & 0.34$_{(0.71)}$\\
 SCAD & 2.54$_{(2.73)}$ & 1.41$_{(1.46)}$ & 0.08$_{(0.02)}$ & 0.82$_{(0.77)}$ & 0.05$_{(0.01)}$ & 0.69$_{(0.29)}$ & 0.64$_{(0.80)}$ & 0.05$_{(0.06)}$ & 2.52$_{(2.97)}$ & 0.94$_{(1.61)}$ & 0.00$_{(0.00)}$ & 0.03$_{(0.00)}$ & 0.02$_{(0.00)}$ & 0.39$_{(0.08)}$ & 0.67$_{(0.97)}$ & 0.27$_{(0.66)}$\\
 MCP & 2.54$_{(2.73)}$ & 1.41$_{(1.46)}$ & 0.08$_{(0.02)}$ & 0.82$_{(0.77)}$ & 0.05$_{(0.01)}$ & 0.69$_{(0.29)}$ & 0.64$_{(0.80)}$ & 0.05$_{(0.06)}$ & 2.51$_{(2.97)}$ & 0.94$_{(1.61)}$ & 0.00$_{(0.00)}$ & 0.03$_{(0.00)}$ & 0.02$_{(0.00)}$ & 0.39$_{(0.08)}$ & 0.67$_{(0.97)}$ & 0.27$_{(0.66)}$\\
 \hline
 Cont 3\\
 LASSO & 2.30$_{(2.65)}$ & 1.25$_{(1.36)}$ & 0.06$_{(0.01)}$ & 0.63$_{(0.45)}$ & 0.14$_{(0.02)}$ & 1.79$_{(0.75)}$ & 0.47$_{(0.75)}$ & 0.08$_{(0.17)}$ & 2.64$_{(2.91)}$ & 1.05$_{(1.66)}$ & 0.00$_{(0.00)}$ & 0.08$_{(0.01)}$ & 0.02$_{(0.00)}$ & 0.34$_{(0.07)}$ & 0.73$_{(0.94)}$ & 0.34$_{(0.71)}$\\
 SCAD & 2.27$_{(2.64)}$ & 1.22$_{(1.32)}$ & 0.06$_{(0.01)}$ & 0.62$_{(0.46)}$ & 0.16$_{(0.01)}$ & 1.82$_{(0.77)}$ & 0.44$_{(0.75)}$ & 0.08$_{(0.16)}$ & 2.47$_{(2.89)}$ & 0.99$_{(1.61)}$ & 0.01$_{(0.00)}$ & 0.07$_{(0.01)}$ & 0.02$_{(0.00)}$ & 0.36$_{(0.08)}$ & 0.66$_{(0.94)}$ & 0.3$_{(0.68)}$\\
 MCP & 2.27$_{(2.64)}$ & 1.22$_{(1.32)}$ & 0.06$_{(0.01)}$ & 0.62$_{(0.46)}$ & 0.16$_{(0.01)}$ & 1.82$_{(0.76)}$ & 0.44$_{(0.75)}$ & 0.08$_{(0.15)}$ & 2.48$_{(2.88)}$ & 0.99$_{(1.60)}$ & 0.01$_{(0.00)}$ & 0.07$_{(0.01)}$ & 0.02$_{(0.00)}$ & 0.36$_{(0.08)}$ & 0.66$_{(0.94)}$ & 0.30$_{(0.68)}$\\
 \hline
\end{tabular}
\end{center}
\end{sidewaystable}

\begin{sidewaystable}[htb]
\renewcommand{\arraystretch}{1.5}
\caption{\small  Variable selection results for Model 2 with $n=100_{(n=200)}$, $\pi_1=0.5$, and $\rho_{ij}=0$.} \label{Table3}
\begin{center}
\scriptsize
\setlength{\tabcolsep}{1mm}
\begin{tabular}{c  cc| cc|cc| cc|   cc| cc|cc| cc}
\hline\hline
        &\multicolumn{8}{c}{FMR}      &\multicolumn{8}{|c}{trim}  \\
\cline{2-9}\cline{10-17}
 &\multicolumn{2}{c}{Correct } &\multicolumn{2}{c}{Incorrect} &\multicolumn{2}{c}{MME } &\multicolumn{2}{c}{Model Accuracy} &\multicolumn{2}{|c}{Correct } &\multicolumn{2}{c}{Incorrect}    &\multicolumn{2}{c}{MME } &\multicolumn{2}{c}{Model Accuracy}    \\
\cline{2-3}\cline{4-5}\cline{6-7}\cline{8-9}\cline{10-11}\cline{12-13}\cline{14-15}\cline{16-17}
 Method  &Comp1(2)&Comp2(2)&Comp1&Comp2 &Comp1&Comp2 &Comp1&Comp2 &Comp1(2)&Comp2(2)&Comp1&Comp2 &Comp1&Comp2 &Comp1&Comp2\\
\hline
 Cont 1\\
 LASSO & 1.45$_{(1.62)}$ & 1.62$_{(1.77)}$ & 0.33$_{(0.29)}$ & 1.06$_{(1.03)}$ & 0.38$_{(0.12)}$ & 0.17$_{(0.11)}$ & 0.42$_{(0.61)}$ & 0.02$_{(0.02)}$ & 1.36$_{(1.83)}$ & 1.41$_{(1.85)}$ & 0.15$_{(0.02)}$ & 0.09$_{(0.01)}$ & 0.20$_{(0.04)}$ & 0.17$_{(0.05)}$ & 0.43$_{(0.82)}$ & 0.46$_{(0.85)}$\\
 SCAD & 1.42$_{(1.59)}$ & 1.61$_{(1.74)}$ & 0.34$_{(0.29)}$ & 1.05$_{(1.02)}$ & 0.40$_{(0.11)}$ & 0.19$_{(0.11)}$ & 0.4$_{(0.59)}$ & 0.02$_{(0.02)}$ & 1.26$_{(1.78)}$ & 1.25$_{(1.76)}$ & 0.12$_{(0.02)}$ & 0.09$_{(0.01)}$ & 0.23$_{(0.04)}$ & 0.19$_{(0.06)}$ & 0.37$_{(0.78)}$ & 0.38$_{(0.78)}$\\
 MCP & 1.42$_{(1.59)}$ & 1.61$_{(1.73)}$ & 0.34$_{(0.29)}$ & 1.05$_{(1.02)}$ & 0.40$_{(0.11)}$ & 0.19$_{(0.11)}$ & 0.4$_{(0.59)}$ & 0.02$_{(0.02)}$ & 1.26$_{(1.78)}$ & 1.26$_{(1.75)}$ & 0.11$_{(0.02)}$ & 0.08$_{(0.01)}$ & 0.23$_{(0.04)}$ & 0.19$_{(0.06)}$ & 0.37$_{(0.78)}$ & 0.38$_{(0.78)}$\\
 \hline
 Cont 2\\
 LASSO & 1.21$_{(1.04)}$ & 1.42$_{(1.36)}$ & 0.76$_{(0.66)}$ & 1.20$_{(1.19)}$ & 11.97$_{(7.87)}$ & 0.54$_{(0.49)}$ & 0.13$_{(0.12)}$ & 0.01$_{(0.00)}$ & 1.44$_{(1.81)}$ & 1.44$_{(1.84)}$ & 0.11$_{(0.04)}$ & 0.08$_{(0.02)}$ & 0.18$_{(0.04)}$ & 0.20$_{(0.04)}$ & 0.53$_{(0.80)}$ & 0.54$_{(0.84)}$\\
 SCAD & 1.19$_{(1.00)}$ & 1.39$_{(1.32)}$ & 0.71$_{(0.64)}$ & 1.16$_{(1.15)}$ & 11.86$_{(8.28)}$ & 0.55$_{(0.47)}$ & 0.12$_{(0.11)}$ & 0.02$_{(0.00)}$ & 1.35$_{(1.76)}$ & 1.27$_{(1.79)}$ & 0.11$_{(0.03)}$ & 0.06$_{(0.01)}$ & 0.19$_{(0.04)}$ & 0.25$_{(0.04)}$ & 0.46$_{(0.76)}$ & 0.43$_{(0.79)}$\\
 MCP & 1.19$_{(1.00)}$ & 1.39$_{(1.32)}$ & 0.71$_{(0.64)}$ & 1.16$_{(1.15)}$ & 11.86$_{(8.28)}$ & 0.55$_{(0.47)}$ & 0.12$_{(0.11)}$ & 0.02$_{(0.00)}$ & 1.36$_{(1.77)}$ & 1.28$_{(1.79)}$ & 0.10$_{(0.03)}$ & 0.06$_{(0.01)}$ & 0.19$_{(0.04)}$ & 0.25$_{(0.04)}$ & 0.46$_{(0.76)}$ & 0.43$_{(0.79)}$\\
 \hline
 Cont 3\\
 LASSO & 1.00$_{(1.09)}$ & 1.44$_{(1.41)}$ & 0.67$_{(0.52)}$ & 1.18$_{(1.22)}$ & 9.92$_{(3.10)}$ & 0.53$_{(0.48)}$ & 0.14$_{(0.17)}$ & 0.01$_{(0.01)}$ & 0.90$_{(1.47)}$ & 1.34$_{(1.74)}$ & 0.14$_{(0.07)}$ & 0.18$_{(0.08)}$ & 0.53$_{(0.06)}$ & 0.28$_{(0.06)}$ & 0.28$_{(0.65)}$ & 0.40$_{(0.75)}$\\
 SCAD & 0.96$_{(1.04)}$ & 1.41$_{(1.37)}$ & 0.62$_{(0.51)}$ & 1.17$_{(1.20)}$ & 10.70$_{(3.29)}$ & 0.51$_{(0.47)}$ & 0.14$_{(0.17)}$ & 0.01$_{(0.00)}$ & 0.77$_{(1.24)}$ & 1.22$_{(1.63)}$ & 0.13$_{(0.04)}$ & 0.16$_{(0.11)}$ & 0.52$_{(0.10)}$ & 0.35$_{(0.08)}$ & 0.21$_{(0.54)}$ & 0.34$_{(0.63)}$\\
 MCP & 0.96$_{(1.04)}$ & 1.41$_{(1.37)}$ & 0.62$_{(0.51)}$ & 1.17$_{(1.20)}$ & 10.70$_{(3.29)}$ & 0.51$_{(0.47)}$ & 0.14$_{(0.17)}$ & 0.01$_{(0.00)}$ & 0.77$_{(1.24)}$ & 1.22$_{(1.62)}$ & 0.13$_{(0.04)}$ & 0.16$_{(0.11)}$ & 0.52$_{(0.10)}$ & 0.35$_{(0.08)}$ & 0.21$_{(0.54)}$ & 0.34$_{(0.63)}$\\
\hline
\end{tabular}
\end{center}
\end{sidewaystable}

\begin{sidewaystable}[htb]
\renewcommand{\arraystretch}{1.5}
\caption{\small  Variable selection results for Model 2 with $n=100_{n=200}$, $\pi_1=0.7$, and $\rho_{ij}=0$.} \label{Table4}
\begin{center}
\scriptsize
\setlength{\tabcolsep}{1mm}
\begin{tabular}{c  cc| cc|cc| cc|   cc| cc|cc| cc}
\hline\hline
        &\multicolumn{8}{c}{FMR}      &\multicolumn{8}{|c}{trim}  \\
\cline{2-9}\cline{10-17}
 &\multicolumn{2}{c}{Correct } &\multicolumn{2}{c}{Incorrect} &\multicolumn{2}{c}{MME } &\multicolumn{2}{c}{Model Accuracy} &\multicolumn{2}{|c}{Correct } &\multicolumn{2}{c}{Incorrect}    &\multicolumn{2}{c}{MME } &\multicolumn{2}{c}{Model Accuracy}    \\
\cline{2-3}\cline{4-5}\cline{6-7}\cline{8-9}\cline{10-11}\cline{12-13}\cline{14-15}\cline{16-17}
 Method  &Comp1(2)&Comp2(2)&Comp1&Comp2 &Comp1&Comp2 &Comp1&Comp2 &Comp1(2)&Comp2(2)&Comp1&Comp2 &Comp1&Comp2 &Comp1&Comp2\\
\hline
Cont 1\\
LASSO & 1.63$_{(1.67)}$ & 1.4$_{(1.44)}$ & 0.31$_{(0.24)}$ & 0.91$_{(0.86)}$ & 0.18$_{(0.06)}$ & 0.63$_{(0.3)}$ & 0.57$_{(0.71)}$ & 0.04$_{(0.04)}$ & 1.56$_{(1.96)}$ & 0.98$_{(1.54)}$ & 0.07$_{(0.01)}$ & 0.22$_{(0.12)}$ & 0.11$_{(0.02)}$ & 0.61$_{(0.2)}$ & 0.60$_{(0.95)}$ & 0.17$_{(0.53)}$\\
SCAD & 1.61$_{(1.65)}$ & 1.37$_{(1.42)}$ & 0.23$_{(0.22)}$ & 0.90$_{(0.85)}$ & 0.18$_{(0.06)}$ & 0.63$_{(0.38)}$ & 0.59$_{(0.70)}$ & 0.04$_{(0.03)}$ & 1.44$_{(1.87)}$ & 0.84$_{(1.34)}$ & 0.05$_{(0.01)}$ & 0.20$_{(0.11)}$ & 0.12$_{(0.02)}$ & 0.66$_{(0.23)}$ & 0.53$_{(0.89)}$ & 0.12$_{(0.44)}$\\
MCP & 1.61$_{(1.65)}$ & 1.37$_{(1.42)}$ & 0.23$_{(0.22)}$ & 0.90$_{(0.85)}$ & 0.18$_{(0.06)}$ & 0.63$_{(0.38)}$ & 0.59$_{(0.70)}$ & 0.04$_{(0.03)}$ & 1.42$_{(1.88)}$ & 0.85$_{(1.36)}$ & 0.05$_{(0.01)}$ & 0.20$_{(0.10)}$ & 0.12$_{(0.02)}$ & 0.66$_{(0.23)}$ & 0.53$_{(0.89)}$ & 0.12$_{(0.44)}$\\
\hline
Cont 2\\
LASSO & 1.31$_{(1.1)}$ & 1.17$_{(1.09)}$ & 0.63$_{(0.64)}$ & 0.79$_{(0.76)}$ & 9.29$_{(8.06)}$ & 1.15$_{(1.13)}$ & 0.26$_{(0.19)}$ & 0.1$_{(0.03)}$ & 1.43$_{(1.88)}$ & 0.94$_{(1.37)}$ & 0.09$_{(0.01)}$ & 0.32$_{(0.15)}$ & 0.11$_{(0.02)}$ & 0.84$_{(0.21)}$ & 0.56$_{(0.9)}$ & 0.18$_{(0.44)}$\\
SCAD & 1.27$_{(1.07)}$ & 1.16$_{(1.07)}$ & 0.57$_{(0.65)}$ & 0.79$_{(0.75)}$ & 10.62$_{(8.34)}$ & 1.13$_{(1.12)}$ & 0.25$_{(0.17)}$ & 0.09$_{(0.03)}$ & 1.42$_{(1.85)}$ & 0.82$_{(1.29)}$ & 0.04$_{(0.00)}$ & 0.28$_{(0.13)}$ & 0.10$_{(0.02)}$ & 0.93$_{(0.22)}$ & 0.58$_{(0.88)}$ & 0.14$_{(0.42)}$\\
MCP & 1.27$_{(1.07)}$ & 1.16$_{(1.07)}$ & 0.57$_{(0.65)}$ & 0.79$_{(0.75)}$ & 10.62$_{(8.34)}$ & 1.13$_{(1.12)}$ & 0.25$_{(0.17)}$ & 0.09$_{(0.03)}$ & 1.37$_{(1.84)}$ & 0.82$_{(1.28)}$ & 0.04$_{(0.00)}$ & 0.29$_{(0.13)}$ & 0.10$_{(0.02)}$ & 0.93$_{(0.22)}$ & 0.57$_{(0.88)}$ & 0.14$_{(0.42)}$\\
\hline
Cont 3\\
LASSO & 1.07$_{(1.17)}$ & 1.16$_{(1.05)}$ & 0.62$_{(0.47)}$ & 0.76$_{(0.75)}$ & 10.03$_{(1.75)}$ & 1.11$_{(1.12)}$ & 0.14$_{(0.23)}$ & 0.08$_{(0.03)}$ & 0.82$_{(0.99)}$ & 1.02$_{(1.14)}$ & 0.14$_{(0.1)}$ & 0.47$_{(0.5)}$ & 8.37$_{(11.15)}$ & 1.14$_{(1.22)}$ & 0.28$_{(0.41)}$ & 0.15$_{(0.22)}$\\
SCAD & 1.05$_{(1.15)}$ & 1.14$_{(1.04)}$ & 0.63$_{(0.46)}$ & 0.73$_{(0.72)}$ & 10.81$_{(1.86)}$ & 1.09$_{(1.11)}$ & 0.15$_{(0.23)}$ & 0.07$_{(0.03)}$ & 0.78$_{(0.91)}$ & 0.91$_{(1.07)}$ & 0.08$_{(0.05)}$ & 0.44$_{(0.48)}$ & 7.37$_{(15.9)}$ & 1.14$_{(1.24)}$ & 0.28$_{(0.40)}$ & 0.13$_{(0.20)}$\\
MCP & 1.05$_{(1.15)}$ & 1.14$_{(1.04)}$ & 0.63$_{(0.46)}$ & 0.73$_{(0.72)}$ & 10.81$_{(1.86)}$ & 1.09$_{(1.11)}$ & 0.15$_{(0.23)}$ & 0.07$_{(0.03)}$ & 0.78$_{(0.92)}$ & 0.91$_{(1.08)}$ & 0.08$_{(0.05)}$ & 0.44$_{(0.48)}$ & 7.37$_{(15.9)}$ & 1.14$_{(1.24)}$ & 0.28$_{(0.40)}$ & 0.13$_{(0.20)}$\\
\hline
\end{tabular}
\end{center}
\end{sidewaystable}

\begin{sidewaystable}[htb]
\renewcommand{\arraystretch}{1.5}
\caption{\small  Variable selection results for Model 1 with $n=100_{n=200}$, $\pi_1=0.5$, and $\rho_{ij}=0.5^{|i-j|}$.} \label{Table5}
\begin{center}
\scriptsize
\setlength{\tabcolsep}{1mm}
\begin{tabular}{c  cc| cc|cc| cc|   cc| cc|cc| cc}
\hline\hline
        &\multicolumn{8}{c}{FMR}      &\multicolumn{8}{|c}{trim}  \\
\cline{2-9}\cline{10-17}
 &\multicolumn{2}{c}{Correct } &\multicolumn{2}{c}{Incorrect} &\multicolumn{2}{c}{MME } &\multicolumn{2}{c}{Model Accuracy} &\multicolumn{2}{|c}{Correct } &\multicolumn{2}{c}{Incorrect}    &\multicolumn{2}{c}{MME } &\multicolumn{2}{c}{Model Accuracy}    \\
\cline{2-3}\cline{4-5}\cline{6-7}\cline{8-9}\cline{10-11}\cline{12-13}\cline{14-15}\cline{16-17}
 Method  &Comp1(3)&Comp2(2)&Comp1&Comp2 &Comp1&Comp2 &Comp1&Comp2 &Comp1(3)&Comp2(2)&Comp1&Comp2 &Comp1&Comp2 &Comp1&Comp2\\
\hline
Cont 1\\
LASSO & 2.40$_{(2.71)}$ & 1.66$_{(1.82)}$ & 0.02$_{(0.02)}$ & 1.00$_{(1.00)}$ & 0.12$_{(0.03)}$ & 0.11$_{(0.04)}$ & 0.48$_{(0.75)}$ & 0.00$_{(0.00)}$ & 2.29$_{(2.70)}$ & 1.39$_{(1.76)}$ & 0.00$_{(0.00)}$ & 0.01$_{(0.01)}$ & 0.09$_{(0.02)}$ & 0.12$_{(0.05)}$ & 0.48$_{(0.76)}$ & 0.52$_{(0.79)}$ \\
SCAD & 2.37$_{(2.65)}$ & 1.62$_{(1.78)}$ & 0.02$_{(0.02)}$ & 1.00$_{(1.00)}$ & 0.12$_{(0.03)}$ & 0.12$_{(0.05)}$ & 0.46$_{(0.70)}$ & 0.00$_{(0.00)}$ & 2.11$_{(2.56)}$ & 1.27$_{(1.67)}$ & 0.00$_{(0.00)}$ & 0.00$_{(0.00)}$ & 0.13$_{(0.03)}$ & 0.13$_{(0.06)}$ & 0.37$_{(0.68)}$ & 0.47$_{(0.71)}$ \\
MCP & 2.37$_{(2.65)}$ & 1.62$_{(1.78)}$ & 0.02$_{(0.02)}$ & 1.00$_{(1.00)}$ & 0.12$_{(0.03)}$ & 0.12$_{(0.05)}$ & 0.46$_{(0.70)}$ & 0.00$_{(0.00)}$ & 2.12$_{(2.56)}$ & 1.26$_{(1.66)}$ & 0.00$_{(0.00)}$ & 0.00$_{(0.00)}$ & 0.13$_{(0.03)}$ & 0.13$_{(0.06)}$ & 0.37$_{(0.68)}$ & 0.47$_{(0.71)}$ \\
\hline
Cont 2\\
LASSO & 1.92$_{(2.04)}$ & 1.50$_{(1.70)}$ & 0.09$_{(0.06)}$ & 0.97$_{(0.98)}$ & 0.50$_{(0.21)}$ & 0.20$_{(0.06)}$ & 0.29$_{(0.40)}$ & 0.02$_{(0.01)}$ & 2.23$_{(2.7)}$ & 1.39$_{(1.77)}$ & 0.01$_{(0.00)}$ & 0.01$_{(0.00)}$ & 0.09$_{(0.02)}$ & 0.13$_{(0.04)}$ & 0.47$_{(0.76)}$ & 0.50$_{(0.81)}$\\
SCAD & 1.85$_{(1.98)}$ & 1.47$_{(1.68)}$ & 0.10$_{(0.07)}$ & 0.97$_{(0.98)}$ & 0.52$_{(0.22)}$ & 0.22$_{(0.06)}$ & 0.24$_{(0.35)}$ & 0.02$_{(0.01)}$ & 2.03$_{(2.61)}$ & 1.18$_{(1.70)}$ & 0.01$_{(0.00)}$ & 0.01$_{(0.00)}$ & 0.12$_{(0.03)}$ & 0.15$_{(0.04)}$ & 0.36$_{(0.69)}$ & 0.41$_{(0.76)}$\\
MCP & 1.85$_{(1.98)}$ & 1.47$_{(1.68)}$ & 0.10$_{(0.07)}$ & 0.97$_{(0.98)}$ & 0.52$_{(0.22)}$ & 0.22$_{(0.06)}$ & 0.24$_{(0.35)}$ & 0.02$_{(0.01)}$ & 2.04$_{(2.61)}$ & 1.19$_{(1.70)}$ & 0.01$_{(0.00)}$ & 0.01$_{(0.00)}$ & 0.12$_{(0.03)}$ & 0.15$_{(0.04)}$ & 0.36$_{(0.69)}$ & 0.41$_{(0.76)}$ \\
\hline
Cont 3\\
LASSO & 1.56$_{(1.73)}$ & 1.24$_{(1.62)}$ & 0.19$_{(0.10)}$ & 0.85$_{(0.93)}$ & 1.42$_{(0.47)}$ & 0.78$_{(0.09)}$ & 0.13$_{(0.23)}$ & 0.06$_{(0.04)}$ & 1.67$_{(2.61)}$ & 1.10$_{(1.74)}$ & 0.01$_{(0.00)}$ & 0.04$_{(0.01)}$ & 0.17$_{(0.03)}$ & 0.19$_{(0.04)}$ & 0.28$_{(0.74)}$ & 0.38$_{(0.79)}$\\
SCAD & 1.55$_{(1.71)}$ & 1.23$_{(1.58)}$ & 0.19$_{(0.09)}$ & 0.85$_{(0.92)}$ & 1.52$_{(0.45)}$ & 0.80$_{(0.10)}$ & 0.12$_{(0.22)}$ & 0.06$_{(0.04)}$ & 1.45$_{(2.51)}$ & 0.96$_{(1.62)}$ & 0.01$_{(0.00)}$ & 0.02$_{(0.01)}$ & 0.2$_{(0.03)}$ & 0.22$_{(0.04)}$ & 0.22$_{(0.68)}$ & 0.30$_{(0.72)}$\\
MCP & 1.55$_{(1.71)}$ & 1.21$_{(1.58)}$ & 0.19$_{(0.09)}$ & 0.85$_{(0.92)}$ & 1.52$_{(0.45)}$ & 0.80$_{(0.10)}$ & 0.12$_{(0.22)}$ & 0.06$_{(0.04)}$ & 1.40$_{(2.51)}$ & 0.98$_{(1.62)}$ & 0.01$_{(0.00)}$ & 0.02$_{(0.01)}$ & 0.20$_{(0.03)}$ & 0.22$_{(0.04)}$ & 0.22$_{(0.68)}$ & 0.30$_{(0.72)}$ \\
\hline
\end{tabular}
\end{center}
\end{sidewaystable}

\begin{sidewaystable}[htb]
\renewcommand{\arraystretch}{1.5}
\caption{\small  Variable selection results for Model 1 with $n=100_{n=200}$, $\pi_1=0.7$, and $\rho_{ij}=0.5^{|i-j|}$.} \label{Table6}
\begin{center}
\scriptsize
\setlength{\tabcolsep}{1mm}
\begin{tabular}{c  cc| cc|cc| cc|   cc| cc|cc| cc}
\hline\hline
        &\multicolumn{8}{c}{FMR}      &\multicolumn{8}{|c}{trim}  \\
\cline{2-9}\cline{10-17}
 &\multicolumn{2}{c}{Correct } &\multicolumn{2}{c}{Incorrect} &\multicolumn{2}{c}{MME } &\multicolumn{2}{c}{Model Accuracy} &\multicolumn{2}{|c}{Correct } &\multicolumn{2}{c}{Incorrect}    &\multicolumn{2}{c}{MME } &\multicolumn{2}{c}{Model Accuracy}    \\
\cline{2-3}\cline{4-5}\cline{6-7}\cline{8-9}\cline{10-11}\cline{12-13}\cline{14-15}\cline{16-17}
 Method  &Comp1(3)&Comp2(2)&Comp1&Comp2 &Comp1&Comp2 &Comp1&Comp2 &Comp1(3)&Comp2(2)&Comp1&Comp2 &Comp1&Comp2 &Comp1&Comp2\\
\hline
Cont 1\\
LASSO & 2.64$_{(2.83)}$ & 1.46$_{(1.57)}$ & 0.04$_{(0.02)}$ & 0.96$_{(0.96)}$ & 0.05$_{(0.01)}$ & 0.29$_{(0.14)}$ & 0.68$_{(0.87)}$ & 0.01$_{(0.00)}$ & 2.49$_{(2.87)}$ & 0.87$_{(1.38)}$ & 0.00$_{(0.00)}$ & 0.03$_{(0.01)}$ & 0.05$_{(0.01)}$ & 0.37$_{(0.15)}$ & 0.60$_{(0.89)}$ & 0.20$_{(0.52)}$\\
SCAD & 2.58$_{(2.81)}$ & 1.42$_{(1.52)}$ & 0.04$_{(0.02)}$ & 0.95$_{(0.96)}$ & 0.06$_{(0.01)}$ & 0.34$_{(0.17)}$ & 0.62$_{(0.85)}$ & 0.02$_{(0.00)}$ & 2.42$_{(2.80)}$ & 0.76$_{(1.21)}$ & 0.00$_{(0.00)}$ & 0.04$_{(0.01)}$ & 0.06$_{(0.01)}$ & 0.42$_{(0.16)}$ & 0.55$_{(0.83)}$ & 0.15$_{(0.39)}$\\
MCP & 2.58$_{(2.81)}$ & 1.42$_{(1.52)}$ & 0.04$_{(0.02)}$ & 0.95$_{(0.96)}$ & 0.06$_{(0.01)}$ & 0.34$_{(0.17)}$ & 0.62$_{(0.85)}$ & 0.02$_{(0.00)}$ & 2.42$_{(2.80)}$ & 0.76$_{(1.2)}$ & 0.00$_{(0.00)}$ & 0.04$_{(0.01)}$ & 0.06$_{(0.01)}$ & 0.42$_{(0.16)}$ & 0.55$_{(0.83)}$ & 0.15$_{(0.39)}$\\
\hline
Cont 2\\
LASSO & 2.46$_{(2.16)}$ & 1.35$_{(1.22)}$ & 0.03$_{(0.08)}$ & 0.65$_{(0.69)}$ & 0.05$_{(0.24)}$ & 0.53$_{(2.29)}$ & 0.60$_{(0.36)}$ & 0.05$_{(0.06)}$ & 2.71$_{(2.32)}$ & 1.36$_{(0.87)}$ & 0.00$_{(0.00)}$ & 0.04$_{(0.09)}$ & 0.02$_{(0.07)}$ & 0.17$_{(0.49)}$ & 0.81$_{(0.56)}$ & 0.48$_{(0.22)}$\\
SCAD & 2.45$_{(2.11)}$ & 1.34$_{(1.20)}$ & 0.03$_{(0.06)}$ & 0.65$_{(0.69)}$ & 0.06$_{(0.28)}$ & 0.6$_{(2.47)}$ & 0.59$_{(0.33)}$ & 0.05$_{(0.05)}$ & 2.65$_{(2.17)}$ & 1.18$_{(0.64)}$ & 0.00$_{(0.00)}$ & 0.01$_{(0.08)}$ & 0.02$_{(0.08)}$ & 0.18$_{(0.53)}$ & 0.76$_{(0.51)}$ & 0.40$_{(0.15)}$\\
MCP & 2.45$_{(2.11)}$ & 1.34$_{(1.2)}$ & 0.03$_{(0.06)}$ & 0.65$_{(0.69)}$ & 0.06$_{(0.28)}$ & 0.6$_{(2.47)}$ & 0.59$_{(0.33)}$ & 0.05$_{(0.05)}$ & 2.65$_{(2.17)}$ & 1.18$_{(0.64)}$ & 0.00$_{(0.00)}$ & 0.01$_{(0.08)}$ & 0.02$_{(0.08)}$ & 0.18$_{(0.53)}$ & 0.76$_{(0.51)}$ & 0.40$_{(0.15)}$\\
\hline
Cont 3\\
LASSO & 1.97$_{(2.41)}$ & 1.15$_{(1.26)}$ & 0.05$_{(0.01)}$ & 0.49$_{(0.39)}$ & 0.38$_{(0.07)}$ & 5.33$_{(1.30)}$ & 0.29$_{(0.57)}$ & 0.15$_{(0.1)}$ & 1.87$_{(2.66)}$ & 0.69$_{(1.26)}$ & 0.01$_{(0.00)}$ & 0.14$_{(0.04)}$ & 0.13$_{(0.02)}$ & 5.24$_{(0.14)}$ & 0.38$_{(0.79)}$ & 0.18$_{(0.48)}$\\
SCAD & 1.94$_{(2.38)}$ & 1.13$_{(1.24)}$ & 0.05$_{(0.01)}$ & 0.48$_{(0.39)}$ & 0.39$_{(0.07)}$ & 5.52$_{(1.33)}$ & 0.26$_{(0.56)}$ & 0.14$_{(0.11)}$ & 1.71$_{(2.58)}$ & 0.53$_{(1.20)}$ & 0.00$_{(0.00)}$ & 0.10$_{(0.02)}$ & 0.15$_{(0.02)}$ & 5.22$_{(0.15)}$ & 0.32$_{(0.73)}$ & 0.12$_{(0.45)}$\\
MCP & 1.94$_{(2.38)}$ & 1.13$_{(1.24)}$ & 0.05$_{(0.01)}$ & 0.48$_{(0.38)}$ & 0.39$_{(0.07)}$ & 5.52$_{(1.33)}$ & 0.26$_{(0.56)}$ & 0.14$_{(0.11)}$ & 1.72$_{(2.58)}$ & 0.54$_{(1.20)}$ & 0.00$_{(0.00)}$ & 0.10$_{(0.02)}$ & 0.15$_{(0.02)}$ & 5.22$_{(0.15)}$ & 0.32$_{(0.73)}$ & 0.12$_{(0.45)}$\\
\hline
\end{tabular}
\end{center}
\end{sidewaystable}

\begin{sidewaystable}[htb]
\renewcommand{\arraystretch}{1.5}
\caption{\small Variable selection results for Model 2 with $n=100_{n=200}$, $\pi_1=0.5$, and $\rho_{ij}=0.5^{|i-j|}$.} \label{Table7}
\begin{center}
\scriptsize
\setlength{\tabcolsep}{1mm}
\begin{tabular}{c  cc| cc|cc| cc|   cc| cc|cc| cc}
\hline\hline
        &\multicolumn{8}{c}{FMR}      &\multicolumn{8}{|c}{trim}  \\
\cline{2-9}\cline{10-17}
 &\multicolumn{2}{c}{Correct } &\multicolumn{2}{c}{Incorrect} &\multicolumn{2}{c}{MME } &\multicolumn{2}{c}{Model Accuracy} &\multicolumn{2}{|c}{Correct } &\multicolumn{2}{c}{Incorrect}    &\multicolumn{2}{c}{MME } &\multicolumn{2}{c}{Model Accuracy}    \\
\cline{2-3}\cline{4-5}\cline{6-7}\cline{8-9}\cline{10-11}\cline{12-13}\cline{14-15}\cline{16-17}
 Method  &Comp1(2)&Comp2(2)&Comp1&Comp2 &Comp1&Comp2 &Comp1&Comp2 &Comp1(2)&Comp2(2)&Comp1&Comp2 &Comp1&Comp2 &Comp1&Comp2\\
\hline
Cont 1\\
LASSO & 1.29$_{(1.52)}$ & 1.54$_{(1.75)}$ & 0.44$_{(0.32)}$ & 1.14$_{(1.04)}$ & 0.42$_{(0.16)}$ & 0.29$_{(0.10)}$ & 0.33$_{(0.53)}$ & 0.02$_{(0.01)}$ & 1.25$_{(1.70)}$ & 1.33$_{(1.67)}$ & 0.17$_{(0.08)}$ & 0.07$_{(0.02)}$ & 0.18$_{(0.05)}$ & 0.19$_{(0.08)}$ & 0.38$_{(0.72)}$ & 0.46$_{(0.72)}$\\
SCAD & 1.27$_{(1.52)}$ & 1.49$_{(1.7)}$ & 0.42$_{(0.28)}$ & 1.13$_{(1.04)}$ & 0.47$_{(0.16)}$ & 0.31$_{(0.12)}$ & 0.34$_{(0.56)}$ & 0.02$_{(0.01)}$ & 1.05$_{(1.60)}$ & 1.20$_{(1.55)}$ & 0.15$_{(0.05)}$ & 0.08$_{(0.01)}$ & 0.26$_{(0.06)}$ & 0.21$_{(0.08)}$ & 0.29$_{(0.65)}$ & 0.37$_{(0.66)}$\\
MCP & 1.27$_{(1.52)}$ & 1.49$_{(1.7)}$ & 0.42$_{(0.28)}$ & 1.13$_{(1.04)}$ & 0.47$_{(0.16)}$ & 0.31$_{(0.12)}$ & 0.34$_{(0.56)}$ & 0.02$_{(0.01)}$ & 1.05$_{(1.59)}$ & 1.19$_{(1.55)}$ & 0.15$_{(0.05)}$ & 0.08$_{(0.01)}$ & 0.26$_{(0.06)}$ & 0.21$_{(0.08)}$ & 0.29$_{(0.65)}$ & 0.37$_{(0.66)}$\\
\hline
Cont 2\\
LASSO & 1.05$_{(1.08)}$ & 1.33$_{(1.39)}$ & 0.74$_{(0.65)}$ & 1.19$_{(1.22)}$ & 16.22$_{(11.33)}$ & 0.57$_{(0.56)}$ & 0.14$_{(0.19)}$ & 0.01$_{(0.01)}$ & 1.20$_{(1.74)}$ & 1.22$_{(1.75)}$ & 0.14$_{(0.03)}$ & 0.12$_{(0.02)}$ & 0.20$_{(0.05)}$ & 0.20$_{(0.06)}$ & 0.32$_{(0.76)}$ & 0.38$_{(0.75)}$\\
SCAD & 1.03$_{(1.09)}$ & 1.31$_{(1.34)}$ & 0.69$_{(0.64)}$ & 1.19$_{(1.19)}$ & 18.98$_{(13.22)}$ & 0.56$_{(0.53)}$ & 0.17$_{(0.19)}$ & 0.02$_{(0.01)}$ & 1.03$_{(1.65)}$ & 1.01$_{(1.63)}$ & 0.12$_{(0.01)}$ & 0.14$_{(0.01)}$ & 0.26$_{(0.06)}$ & 0.29$_{(0.06)}$ & 0.25$_{(0.72)}$ & 0.28$_{(0.66)}$\\
MCP & 1.03$_{(1.09)}$ & 1.31$_{(1.34)}$ & 0.69$_{(0.64)}$ & 1.19$_{(1.19)}$ & 18.98$_{(13.22)}$ & 0.56$_{(0.53)}$ & 0.17$_{(0.19)}$ & 0.02$_{(0.01)}$ & 1.02$_{(1.66)}$ & 1.01$_{(1.63)}$ & 0.12$_{(0.01)}$ & 0.14$_{(0.01)}$ & 0.26$_{(0.06)}$ & 0.29$_{(0.06)}$ & 0.25$_{(0.72)}$ & 0.28$_{(0.66)}$\\
\hline
Cont 3\\
LASSO & 1.02$_{(0.96)}$ & 1.40$_{(1.39)}$ & 0.65$_{(0.63)}$ & 1.26$_{(1.22)}$ & 14.01$_{(5.58)}$ & 0.58$_{(0.60)}$ & 0.15$_{(0.11)}$ & 0.02$_{(0.01)}$ & 0.87$_{(1.28)}$ & 1.17$_{(1.58)}$ & 0.22$_{(0.08)}$ & 0.24$_{(0.12)}$ & 0.95$_{(0.11)}$ & 0.42$_{(0.09)}$ & 0.24$_{(0.48)}$ & 0.32$_{(0.61)}$\\
SCAD & 1.02$_{(0.96)}$ & 1.34$_{(1.37)}$ & 0.63$_{(0.61)}$ & 1.24$_{(1.22)}$ & 14.65$_{(5.86)}$ & 0.58$_{(0.58)}$ & 0.17$_{(0.11)}$ & 0.02$_{(0.01)}$ & 0.69$_{(1.15)}$ & 0.98$_{(1.38)}$ & 0.17$_{(0.09)}$ & 0.21$_{(0.11)}$ & 0.76$_{(0.14)}$ & 0.53$_{(0.12)}$ & 0.19$_{(0.40)}$ & 0.26$_{(0.50)}$\\
MCP & 1.02$_{(0.96)}$ & 1.34$_{(1.37)}$ & 0.63$_{(0.61)}$ & 1.24$_{(1.22)}$ & 14.65$_{(5.86)}$ & 0.58$_{(0.58)}$ & 0.17$_{(0.11)}$ & 0.02$_{(0.01)}$ & 0.70$_{(1.15)}$ & 0.98$_{(1.39)}$ & 0.17$_{(0.09)}$ & 0.21$_{(0.10)}$ & 0.76$_{(0.14)}$ & 0.53$_{(0.12)}$ & 0.19$_{(0.40)}$ & 0.26$_{(0.50)}$\\
\hline
\end{tabular}
\end{center}
\end{sidewaystable}

\begin{sidewaystable}[htb]
\renewcommand{\arraystretch}{1.5}
\caption{\small  Variable selection results for Model 2 with $n=100_{n=200}$, $\pi_1=0.7$, and $\rho_{ij}=0.5^{|i-j|}$.} \label{Table8}
\begin{center}
\scriptsize
\setlength{\tabcolsep}{1mm}
\begin{tabular}{c  cc| cc|cc| cc|   cc| cc|cc| cc}
\hline\hline
        &\multicolumn{8}{c}{FMR}      &\multicolumn{8}{|c}{trim}  \\
\cline{2-9}\cline{10-17}
 &\multicolumn{2}{c}{Correct } &\multicolumn{2}{c}{Incorrect} &\multicolumn{2}{c}{MME } &\multicolumn{2}{c}{Model Accuracy} &\multicolumn{2}{|c}{Correct } &\multicolumn{2}{c}{Incorrect}    &\multicolumn{2}{c}{MME } &\multicolumn{2}{c}{Model Accuracy}    \\
\cline{2-3}\cline{4-5}\cline{6-7}\cline{8-9}\cline{10-11}\cline{12-13}\cline{14-15}\cline{16-17}
 Method  &Comp1(2)&Comp2(2)&Comp1&Comp2 &Comp1&Comp2 &Comp1&Comp2 &Comp1(2)&Comp2(2)&Comp1&Comp2 &Comp1&Comp2 &Comp1&Comp2\\
\hline
Cont 1\\
LASSO & 1.49$_{(1.58)}$ & 1.32$_{(1.48)}$ & 0.31$_{(0.31)}$ & 0.97$_{(0.95)}$ & 0.20$_{(0.10)}$ & 0.93$_{(0.33)}$ & 0.50$_{(0.60)}$ & 0.06$_{(0.03)}$ & 1.37$_{(1.91)}$ & 0.89$_{(1.3)}$ & 0.08$_{(0.01)}$ & 0.28$_{(0.12)}$ & 0.11$_{(0.03)}$ & 0.78$_{(0.22)}$ & 0.50$_{(0.91)}$ & 0.16$_{(0.39)}$\\
SCAD & 1.41$_{(1.54)}$ & 1.26$_{(1.44)}$ & 0.29$_{(0.25)}$ & 0.96$_{(0.92)}$ & 0.20$_{(0.10)}$ & 0.98$_{(0.36)}$ & 0.48$_{(0.62)}$ & 0.05$_{(0.03)}$ & 1.22$_{(1.80)}$ & 0.63$_{(1.12)}$ & 0.04$_{(0.02)}$ & 0.22$_{(0.11)}$ & 0.13$_{(0.03)}$ & 0.97$_{(0.26)}$ & 0.42$_{(0.83)}$ & 0.10$_{(0.29)}$\\
MCP & 1.41$_{(1.54)}$ & 1.26$_{(1.44)}$ & 0.29$_{(0.25)}$ & 0.96$_{(0.92)}$ & 0.20$_{(0.10)}$ & 0.98$_{(0.36)}$ & 0.48$_{(0.62)}$ & 0.05$_{(0.03)}$ & 1.21$_{(1.80)}$ & 0.62$_{(1.11)}$ & 0.04$_{(0.02)}$ & 0.22$_{(0.11)}$ & 0.13$_{(0.03)}$ & 0.97$_{(0.26)}$ & 0.42$_{(0.83)}$ & 0.10$_{(0.29)}$\\
\hline
Cont 2\\
LASSO & 1.15$_{(1.2)}$ & 1.12$_{(1.10)}$ & 0.66$_{(0.59)}$ & 0.78$_{(0.79)}$ & 11.95$_{(10.16)}$ & 1.30$_{(1.36)}$ & 0.24$_{(0.27)}$ & 0.05$_{(0.02)}$ & 1.3$_{(1.79)}$ & 0.93$_{(1.31)}$ & 0.12$_{(0.04)}$ & 0.28$_{(0.14)}$ & 0.13$_{(0.02)}$ & 0.96$_{(0.22)}$ & 0.42$_{(0.83)}$ & 0.18$_{(0.42)}$\\
SCAD & 1.14$_{(1.17)}$ & 1.12$_{(1.07)}$ & 0.62$_{(0.59)}$ & 0.76$_{(0.76)}$ & 17.23$_{(10.70)}$ & 1.27$_{(1.35)}$ & 0.25$_{(0.27)}$ & 0.05$_{(0.03)}$ & 1.08$_{(1.68)}$ & 0.7$_{(1.05)}$ & 0.07$_{(0.02)}$ & 0.26$_{(0.13)}$ & 0.17$_{(0.03)}$ & 1.29$_{(0.29)}$ & 0.35$_{(0.78)}$ & 0.14$_{(0.28)}$\\
MCP & 1.14$_{(1.17)}$ & 1.12$_{(1.07)}$ & 0.62$_{(0.59)}$ & 0.76$_{(0.76)}$ & 17.23$_{(10.70)}$ & 1.26$_{(1.35)}$ & 0.25$_{(0.27)}$ & 0.05$_{(0.03)}$ & 1.08$_{(1.66)}$ & 0.72$_{(1.05)}$ & 0.06$_{(0.02)}$ & 0.27$_{(0.14)}$ & 0.17$_{(0.03)}$ & 1.29$_{(0.29)}$ & 0.35$_{(0.78)}$ & 0.14$_{(0.28)}$\\
\hline
Cont 3\\
LASSO & 1.15$_{(1.20)}$ & 1.12$_{(1.1)}$ & 0.66$_{(0.59)}$ & 0.78$_{(0.79)}$ & 11.95$_{(10.16)}$ & 1.30$_{(1.36)}$ & 0.24$_{(0.27)}$ & 0.05$_{(0.02)}$ & 1.3$_{(1.79)}$ & 0.93$_{(1.31)}$ & 0.12$_{(0.04)}$ & 0.28$_{(0.14)}$ & 0.13$_{(0.02)}$ & 0.96$_{(0.22)}$ & 0.42$_{(0.83)}$ & 0.18$_{(0.42)}$\\
SCAD & 1.14$_{(1.17)}$ & 1.12$_{(1.07)}$ & 0.62$_{(0.59)}$ & 0.76$_{(0.76)}$ & 17.23$_{(10.70)}$ & 1.27$_{(1.35)}$ & 0.25$_{(0.27)}$ & 0.05$_{(0.03)}$ & 1.08$_{(1.68)}$ & 0.70$_{(1.05)}$ & 0.07$_{(0.02)}$ & 0.26$_{(0.13)}$ & 0.17$_{(0.03)}$ & 1.29$_{(0.29)}$ & 0.35$_{(0.78)}$ & 0.14$_{(0.28)}$\\
MCP & 1.14$_{(1.17)}$ & 1.12$_{(1.07)}$ & 0.62$_{(0.59)}$ & 0.76$_{(0.76)}$ & 17.23$_{(10.70)}$ & 1.26$_{(1.35)}$ & 0.25$_{(0.27)}$ & 0.05$_{(0.03)}$ & 1.08$_{(1.66)}$ & 0.72$_{(1.05)}$ & 0.06$_{(0.02)}$ & 0.27$_{(0.14)}$ & 0.17$_{(0.03)}$ & 1.29$_{(0.29)}$ & 0.35$_{(0.78)}$ & 0.14$_{(0.28)}$\\
\hline
\end{tabular}
\end{center}
\end{sidewaystable}

\clearpage
\section*{References}
\begin{description}
\item[] Atkinson, A. C. and Riani, M. (2002). Forward search added-variable t-tests and the effect of masked outliers on model selection. \emph{Biometrika}, 89, 939--946.
\item[] Bondell, H, Kong, D., and Wu, Y. (2016). Fully efficient robust estimation, outlier detection, and variable selection via penalized regression. \emph{Statistica Sinica}, preprint.
\item[] Cai, Z., Fan, J., and Li, R. (2000). Efficient estimation and inferences for varying-coefficient models. \emph{Journal of the American Statistical Association}, 95, 888--902.
\item[] Cantoni, E. and Ronchetti, E. (2001). Robust inference for generalized linear models. \emph{Journal of the American Statistical Association}, 96, 1022--1030.
\item[] Fan, J. and Li, R. (2001). Variable selection via non-concave penalized likelihood and its oracle properties. \emph{Journal of the American Statistical Association}, 96, 1348--1360.
\item[] Fan, J. and Zhang, W. (1999). Statistical estimation in varying coefficient models. \emph{Annals of Statistics}, 27, 1491--1518.
\item[] Fan, Y., Qin, G., and Zhu, Z. (2012). Variable selection in robust regression models for longitudinal data. \emph{Journal of Multivariate Analysis}, 109, 156--167.
\item[] Fr$\ddot{u}$hwirth-Schnatter, S.(2001). Markov chain monte carlo estimation of classical and dynamic switching and mixture models. \emph{Journal of American and Statisical Association}, 96, 194--209.
\item[] Green, P. J. and Richardson, S. (2002). Hidden markov models and disease mapping. \emph{Journal of American and Statistical Association}, 97, 1055--1070.
\item[] Huang, M. and Yao, W. (2012). Mixture of regression models with varying mixing proportions: a semiparametric approach. \emph{Journal of the American Statistical Association}, 107, 711--724.
\item[] Hunter, D.R. and Young, D.S. (2012). Semiparametric mixtures of regressions. \emph{Journal of Nonparametric Statistics}, 24, 19--38.
\item[] Jiang, Y. (2016). Robust variable selection for mixture linear regression models. \emph{Hacettepe Journal of Mathematics and Statistics}, 45, 549--559.
\item[] Khalili, A. and Chen, J. (2007). Variable selection in finite mixture of regression models. \emph{Journal of the American Statistical Association}, 102, 1025--2038.
\item[] Khalili, A., Chen, J. and Lin, S. (2011). Feature selection in finite mixture of sparse normal linear models in high-dimensional feature space. \emph{Biometrics}, 12, 156--172.
\item[] Li, M., Xiang, S. and Yao, W. (2016). Robust estimation of the number of components for mixtures of linear regression models. \emph{Computational Statistics}, 31(4), 1539--1555.
\item[] McLachlan, G. J. and Peel, D. (2000). \emph{Finite Mixture Models}, New York: Wiley.
\item[] M$\ddot{u}$ller, C. and Neykov, N. (2003). Breakdown points of trimmed likelihood estimators and related estimators in generalized linear models. \emph{Journal of Statistical Plan \& Inference}, 116, 503--519.
\item[] M$\ddot{u}$ller, W. and Welsh, A.H. (2005). Outlier robust model selection in linear regression. \emph{Journal of the American Statistical Association}, 100, 1297--1310.
\item[] Neykov, N., Filzmoser, P., Dimova, R., Neytchev, P. (2007). Robust fitting of mixtures using the trimmed likelihood estimator. \emph{Computational Statistics \& Data Analysis}, 52, 299--308.
\item[] Ronchetti, E. (1985). Robust model selection in regression. \emph{Statistics and Probability Letters}, 3, 21--23.
\item[] Ronchetti, E., Field, C., and Blanchard, W. (1997). Robust linear model selection by cross-validation. \emph{Journal of the American Statistical Association}, 92, 1017--1023.
\item[] Ronchetti, E. and Staudte, R. G. (1994). A robust version of Mallows's Cp. \emph{Journal of the American Statistical Association}, 89, 550--559.
\item[] Ronchetti, E. and Trojani, F. (2001). Robust inference with GMM estimators. \emph{Journal of Econometrics}, 101, 37--69.
\item[] Shao, J. (1993). Linear models selection by cross-validation. \emph{Journal of the American Statistical Association}, 88, 486--494.
\item[] Skrondal, A. and Rabe-Hesketh, S. (2004). \emph{Generalized Latent Variable Modeling: Multilevel, Longitudinal, and Structural Equation Models}, Boca Raton, FL: Chapman \& Hall/CRC.
\item[] Tibshirani, R. (1996). Regression shrinkage and selection via the LASSO. \emph{Journal of the Royal Statistical Society, Ser. B}, 58, 267-288.
\item[] Wang, P., Puterman, M. L., Cockburn, I., and Le, N. (1996). Mixed Poisson regression models with covariate dependent rates. \emph{Biometrics}, 52, 381--400.
\item[] Wedel, M. and DeSarbo, W. S. (1993). A latent class binomial logit methodology for the analysis of paired comparison data. \emph{Decision Sci}, 24, 1157--1170.
\item[] Wisnowski, J. W., Simpson, J. R., Montgomery, D. C., and Runger, G. C. (2003). Resampling methods for variable selection in robust regression. \emph{Computational Statistics \& Data Analysis}, 43, 341--355.
\item[] Xia, Y., Tong, H., Li, W., Zhu, L.X. (2002). An adaptive estimation of dimension reduction space. \emph{Journal of the Royal Statistical Society: Series B}, 64, 363--410.
\item[] Xiang, S. and Yao, W. (2017). Semiparametric mixtures of regressions with single-index for model based clustering. arXiv:1708.04142.
\item[]  Yang, G., Xiang, S., Yao, W., and Xu, L. (2017). Robust estimation and outlier detection for varying coefficient models via penalized regression. \emph{Computational Statistics}. Submitted.
\item[] Yang, L., Xiang, S. and Yao, W. (2017). Robust fitting of mixtures of factor analyzers using the trimmed likelihood estimator. \emph{Communications in Statistics - Simulation and Computation}, 42(2), 1280--1291
\item[] Yao, W. and Li, R. (2014). A new regression model: modal linear regression. \emph{Scandinavian Journal of Statistics}, 41, 656--671.
\item[] Young, D.S. and Hunter, D.R. (2010). Mixtures of regressions with predictor-dependent mixing proportions. \emph{Computational Statistics \& Data Analysis}, 54, 2253--2266.
\item[] Zhang, C.H. (2010). Nearly unbiased variable selection under minimax concave penalty. \emph{Annals of Statistics}, 38, 894-942.
\item[] Zhang, R., Zhao, W., and Liu, J. (2013). Robust estimation and variable selection for semiparametric partially linear varying coefficient model based on modal regression. \emph{Journal of Nonparametric Statistics}, 25, 523--544.
\end{description}
\end{document}